\documentclass[aps,prl,twocolumn,amsmath,amssymb,superscriptaddress,showpacs,longbibliography]{revtex4-1}

\usepackage{hyperref}
\pdfoutput=1
\usepackage{graphicx}
\usepackage{epsfig}
\usepackage{epstopdf}

\usepackage{amsmath}
\usepackage{bm}
\usepackage{amssymb}
\usepackage{textcomp}
\usepackage{ulem}

\usepackage{import}
\usepackage{color}
\usepackage[english]{babel}
\usepackage{float}
\usepackage{SIunits}

\begin{document}

\title{Distance dependence of the energy transfer rate from a single semiconductor nanostructure to graphene}

\author{Fran\c{c}ois Federspiel}
\affiliation{Institut de Physique et Chimie des Mat\'eriaux de Strasbourg and NIE, UMR 7504, Universit\'e de Strasbourg and CNRS, 23 rue du L\oe{}ss, BP43, 67034 Strasbourg Cedex 2, France}

\author{Guillaume Froehlicher}
\affiliation{Institut de Physique et Chimie des Mat\'eriaux de Strasbourg and NIE, UMR 7504, Universit\'e de Strasbourg and CNRS, 23 rue du L\oe{}ss, BP43, 67034 Strasbourg Cedex 2, France}

\author{Michel Nasilowski}
\affiliation{Laboratoire de Physique et d'Etude des Mat\'eriaux, ESPCI-ParisTech, 
PSL Research University, Sorbonne Universit\'e UPMC Univ Paris 06, CNRS, 10 rue Vauquelin 75005 Paris, France}

\author{Silvia Pedetti}
\affiliation{Laboratoire de Physique et d'Etude des Mat\'eriaux, ESPCI-ParisTech, 
PSL Research University, Sorbonne Universit\'e UPMC Univ Paris 06, CNRS, 10 rue Vauquelin 75005 Paris, France}

\author{Ather Mahmood}
\affiliation{Institut de Physique et Chimie des Mat\'eriaux de Strasbourg and NIE, UMR 7504, Universit\'e de Strasbourg and CNRS, 23 rue du L\oe{}ss, BP43, 67034 Strasbourg Cedex 2, France}

\author{Bernard Doudin}
\affiliation{Institut de Physique et Chimie des Mat\'eriaux de Strasbourg and NIE, UMR 7504, Universit\'e de Strasbourg and CNRS, 23 rue du L\oe{}ss, BP43, 67034 Strasbourg Cedex 2, France}

\author{Serin Park}
\affiliation{Advanced Materials Division, Korea Research Institute of Chemical Technology, Daejeon 305-343, Korea}

\author{Jeong-O Lee}
\affiliation{Advanced Materials Division, Korea Research Institute of Chemical Technology, Daejeon 305-343, Korea}

\author{David Halley}
\affiliation{Institut de Physique et Chimie des Mat\'eriaux de Strasbourg and NIE, UMR 7504, Universit\'e de Strasbourg and CNRS, 23 rue du L\oe{}ss, BP43, 67034 Strasbourg Cedex 2, France}

\author{Beno\^{i}t Dubertret}
\affiliation{Laboratoire de Physique et d'Etude des Mat\'eriaux, ESPCI-ParisTech, 
PSL Research University, Sorbonne Universit\'e UPMC Univ Paris 06, CNRS, 10 rue Vauquelin 75005 Paris, France}

\author{Pierre Gilliot}
\affiliation{Institut de Physique et Chimie des Mat\'eriaux de Strasbourg and NIE, UMR 7504, Universit\'e de Strasbourg and CNRS, 23 rue du L\oe{}ss, BP43, 67034 Strasbourg Cedex 2, France}

\author{St\'ephane Berciaud}
\email{stephane.berciaud@ipcms.unistra.fr}
\affiliation{Institut de Physique et Chimie des Mat\'eriaux de Strasbourg and NIE, UMR 7504, Universit\'e de Strasbourg and CNRS, 23 rue du L\oe{}ss, BP43, 67034 Strasbourg Cedex 2, France}



\begin{abstract}
The near-field Coulomb interaction between a nano-emitter and a graphene monolayer results in strong F\"orster-type resonant energy transfer and subsequent fluorescence quenching. Here, we investigate the distance dependence of the energy transfer rate from individual, i) zero-dimensional CdSe/CdS nanocrystals and ii) two-dimensional CdSe/CdS/ZnS nanoplatelets to a graphene monolayer. For increasing distances $d$, the energy transfer rate from individual nanocrystals to graphene decays as $1/d^4$. In contrast, the distance dependence of the energy transfer rate from a two-dimensional nanoplatelet to graphene deviates from a simple power law, but is well described by a theoretical model, which considers a thermal distribution of free excitons in a two-dimensional quantum well. Our results show that accurate distance measurements can be performed at the single particle level using graphene-based molecular rulers and that energy transfer allows probing dimensionality effects at the nanoscale.

\end{abstract}

\maketitle

\paragraph{\textbf{Introduction}}
Graphene and colloidal semiconductor nanostructures are model low-dimensional systems, which hold promise for opto-electronic applications~\cite{bonaccorso2010,koppens2014,talapin2010}.
On the one hand, graphene, as a quasi-transparent semi-metal~\cite{nair2008,mak2008} with excellent transport properties~\cite{dassarma2011}, can be seen as an ultimate transparent electrode~\cite{wang2008,kim2009b}. On the other hand, CSNs, in the form of zero-dimensional nanocrystals~\cite{KlimovBOOK} (NCs or quantum dots), one-dimensional quantum rods~\cite{peng2000} and two dimensional nanoplatelets~\cite{ithurria2008,ithurria2011} (NPs, or quantum wells), are very efficient broadband light harvesting systems and size tunable nano-emitters, which are intensively used in a new generation of  light emitting diodes, solar cells and photovoltaic devices~\cite{talapin2010}.

\begin{figure*}[thb]
\begin{center}
\includegraphics[scale=0.6]{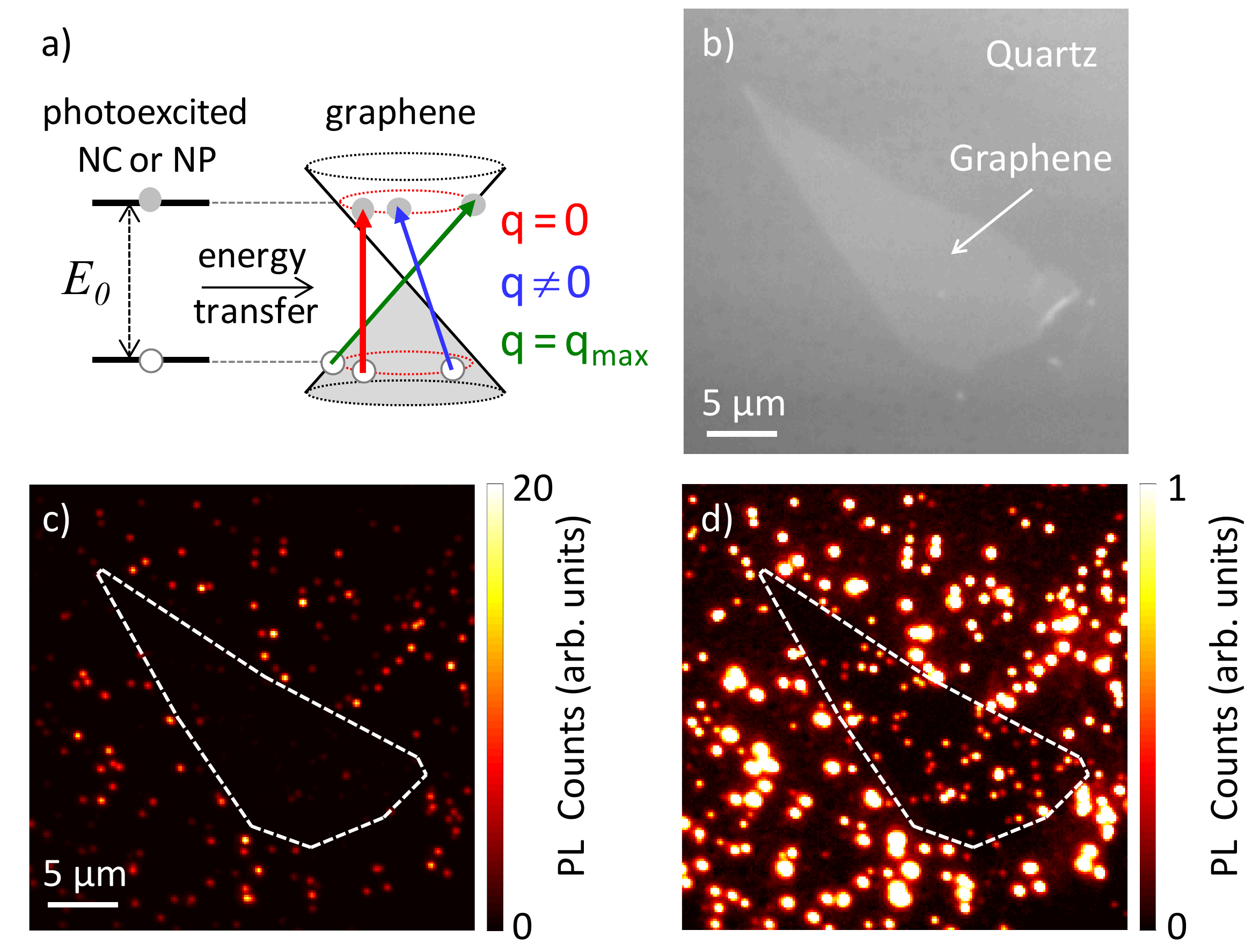}
\caption{a) Schematic representation of the resonant energy transfer process between a photoexcited nanoemitter and undoped graphene. Electronic excitations in graphene with various transferred momenta $q$ are shown with colored arrows.  b) Optical image of an exfoliated graphene sample deposited on fused quartz. c-d) Photoluminescence (PL) image of the same sample, covered with CdSe/CdS/ZnS nanoplatelets, shown with two different linear scales of PL intensity.}
\label{Fig01}
\end{center}
\end{figure*}

There is a growing interest in combining graphene and colloidal semiconductor nanostructures in the form of hybrid systems~\cite{chen2010,ajayi2014,rogez2014} and devices~\cite{konstantatos2012,sun2012,klekachev2012,klekachev2013} with new functionalities and potentially enhanced opto-electronic properties.  The photoresponse  of the graphene NC-hybrid system is governed by interface and  short-range phenomena, such as charge transfer and F\"orster-type resonant energy transfer~\cite{forster1948} (RET) (see Figure~\ref{Fig01}). While photo-induced charge transfer may result in a photogating effect and improved photogain~\cite{konstantatos2012,sun2012}, energy transfer from a photoexcited colloidal semiconductor nanostructure (donor) to a graphene layer (acceptor) may efficiently generate electron-hole pairs in graphene, which is of interest for optoelectronics~\cite{koppens2014}. Importantly, graphene stands out as a uniquely tunable acceptor system, in which distinct regimes of RET can be observed by varying its Fermi level~\cite{velizhanin2011,gomez2011,tielrooij2014,lee2014}. 

Highly efficient RET from individual CdSe/ZnS NCs to graphene, resulting in a quenching of the luminescence signal by more than one order of magnitude, has recently been reported~\cite{chen2010}. Related effects have been observed using other types of semiconductor nanostructures~\cite{ajayi2014,rogez2014,shafran2010,jander2011,lin2013}, fluorescent molecules~\cite{treossi2009,kim2009,gaudreau2013} or NV centers~\cite{stohr2012,tisler2013}. The observation of robust and efficient RET to graphene has stimulated numerous applications in biosensing~\cite{wang2011} and holds promise for distance sensing~\cite{mazzamuto2014} and photodetection. The case of a single colloidal semiconductor nanostructure near a single layer of graphene is of particular interest, since it provides a well-defined and technologically relevant system, in which the sensitivity of the RET rate to the local environment and its distance dependence can be assessed with accuracy. In addition, RET is known to be strongly affected by exciton dimensionality and exciton localization~\cite{halivni2012,rindermann2011}. Colloidal semiconductor nanostructures offer natural ways to explore such effects. 

Here, we investigate RET from i) individual core/shell CdSe/CdS NCs and ii) core/shell CdSe/CdS/ZnS NPs to a graphene monolayer. Using molecular beam epitaxy, we are able to deposit ultrasmooth dielectric spacers of magnesium oxide (MgO), with variable thickness, between graphene and the nanoemitters~\cite{godel2013}. The scaling of the RET rate with the distance $d$ separating graphene from the nanoemitters is then quantitatively determined from the luminescence decays recorded on a collection of individual emitters. In the case of zero-dimensional NCs, the RET rate scales as $1/d^4$, as expected theoretically~\cite{kuhn1970,chance1978,swathi2009,gomez2011,velizhanin2011,gaudreau2013}. Interestingly, although the RET rate of individual two-dimensional NPs adsorbed on bare graphene is similar to that observed with zero-dimensional NCs, we find that the RET rate decays less rapidly with increasing distance. Such a behavior is discussed within the framework of energy transfer from free two-dimensional excitons~\cite{basko2000,kos2005} to a two-dimensional acceptor.

\paragraph{\textbf{Methods}}

We have investigated individual colloidal semiconductor nanostructures near a graphene monolayer at room temperature, using a home-built micro-photoluminescence (PL) setup equipped with a $100\times$ $(\rm{NA}=0.90)$ air objective. Core/shell CdSe/CdS NCs ($(9.5\pm1.5)~\rm nm$ in diameter, with peak emission at $\lambda_0=580\rm~nm$, \textit{i.e.,} photon energy of $2.14~\rm eV$), coated with oleylamine and oleic acid ligands, and core/shell/shell CdSe/CdS/ZnS NPs (4 monolayer thick core, $\sim 1.3~\rm nm$ thick shell, $\sim 9\rm~nm$ width, $\sim 22\rm ~nm$ length and peak emission at $\lambda_0=645\rm~nm$, \textit{i.e.,} photon energy of $1.92~\rm eV$), coated with oleate ligands, were synthesized following previous works~\cite{mohamed2005,mahler2008,ithurria2011,mahler2012,tessier2013} (see also Supporting Information). Using core/shell structures dramatically reduces the possibility of irreversible photoinduced NC or NP ionization and subsequent charge transfer to graphene~\cite{hines1996,tessier2013}. NCs and NPs, dispersed at very low concentration into a $90\%/10\%$ hexane/octane mixture were then dropcast onto graphene samples.

\begin{figure*}[hbt]
\begin{center}
\includegraphics[scale=0.6]{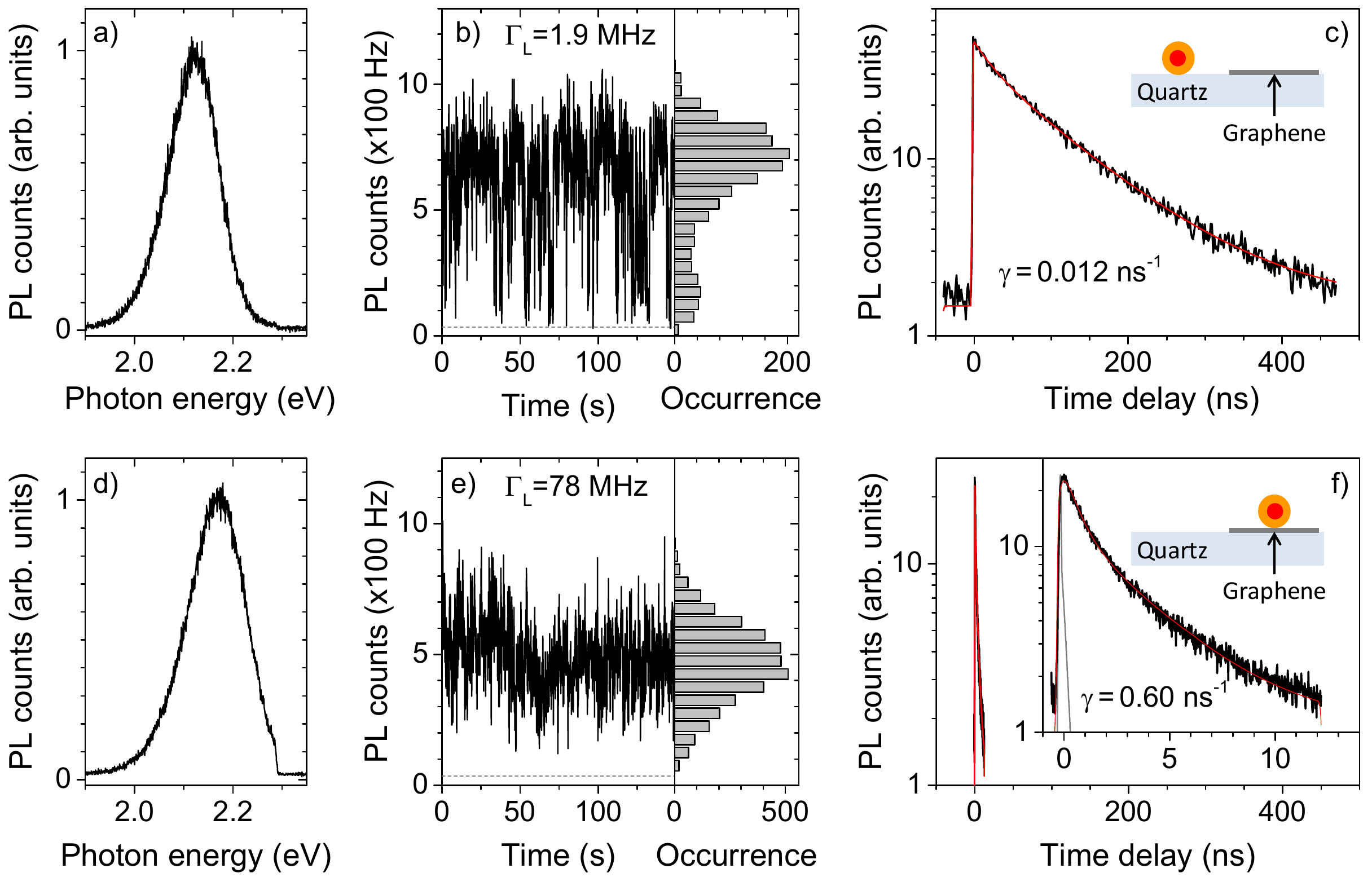}
\caption{ a-c) Photoluminescence (PL) spectrum, PL time trace and PL decay of individual CdSe/CdS nanocrystal deposited on fused quartz. d-f) Same measurements recorded on an another individual CdSe/CdS nanocrystal deposited on a graphene monolayer. All data were collected using a pulsed laser excitation, with a repetition rate $\Gamma_L^{}$ indicated in panels b) and e). The red lines in c) and f) are fits based on bi-exponential functions convoluted with the instrument response function (displayed as a gray line in f).}
\label{Fig02}
\end{center}
\end{figure*}

\begin{figure*}[hbt]
\begin{center}
\includegraphics[scale=0.6]{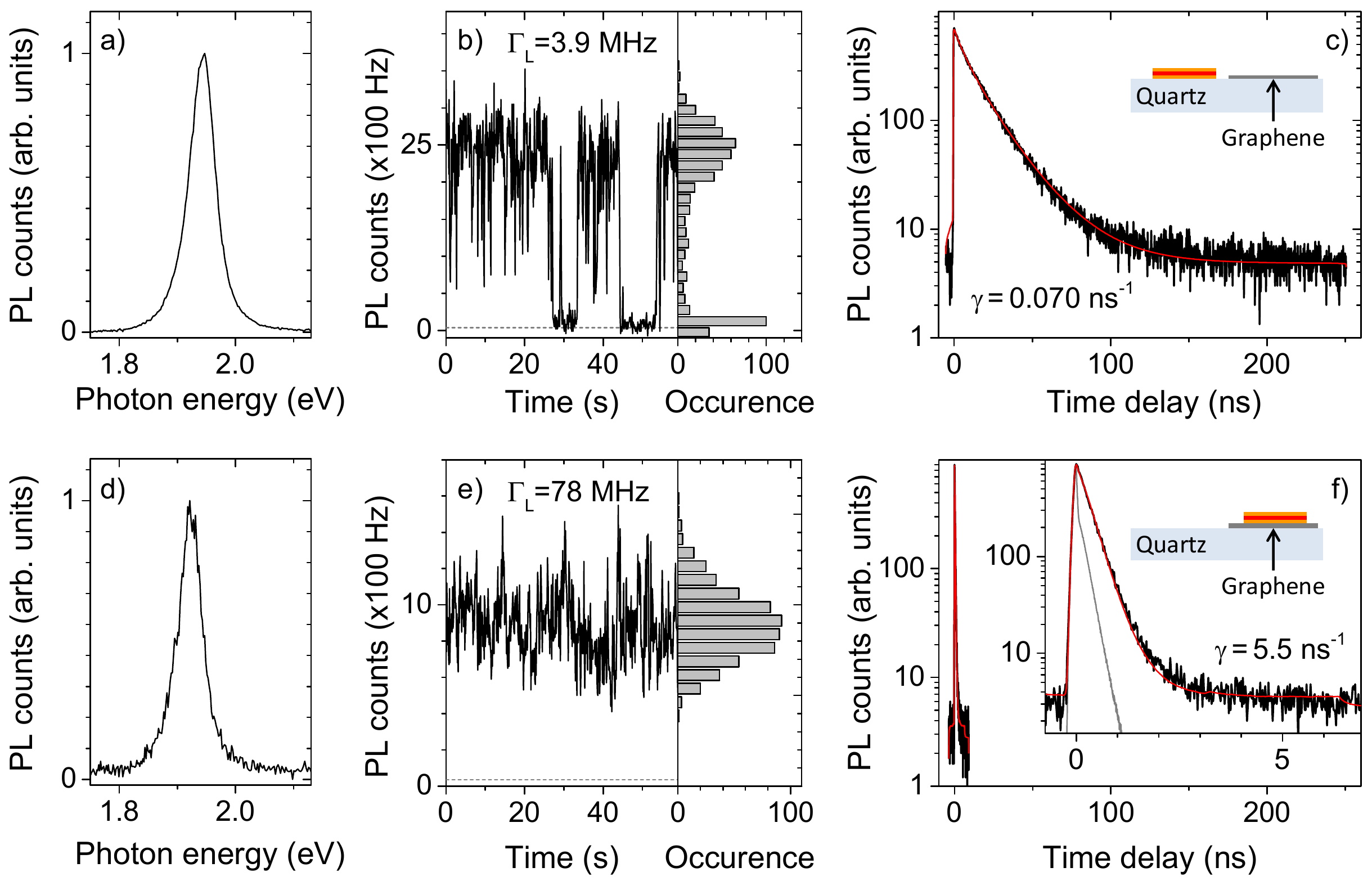}
\caption{ a-c) Photoluminescence (PL) spectrum, PL time trace and PL decay of an individual CdSe/CdS/ZnS nanoplatelet deposited on fused quartz. d-f) Same measurements on another individual CdSe/CdS/ZnS nanoplatelet deposited on a graphene monolayer. All data were collected using a pulsed laser excitation, with a repetition rate $\Gamma_L^{}$ indicated in panels b) and e). The red lines in c) and f) are fits based on stretched exponential functions convoluted with the instrument response function (displayed as a gray line in f).}
\label{Fig03}
\end{center}
\end{figure*}

Measurements on bare graphene (see Figures~\ref{Fig01}--\ref{Fig03}) were performed using mechanically exfoliated graphene monolayers deposited on transparent fused quartz substrates.	The distance dependence of the energy transfer rate was investigated using large area graphene monolayers, grown by low-pressure chemical vapor deposition (LPCVD) on a copper foil, then transfered onto a fused quartz substrate using standard methods~\cite{li2009} (see Supporting Information). In order to vary the distance between graphene and the nano emitters, thin films of MgO were grown on top of graphene in a staircase fashion in a molecular beam epitaxy (MBE) chamber~\cite{godel2013} before deposition of NCs or NPs. We checked by atomic force microscopy that the roughness of the MgO layer is on the order of $0.5~\rm nm$~\cite{wang2008b}. 

The bare graphene and graphene/MgO samples were characterized using a home-built micro-Raman setup (see Supporting Information). Although residual doping on the order of a few $10^{12}~\rm cm^{-2}$ could be observed, the resulting Fermi level shifts relative to the Dirac point are about one order of magnitude smaller than the energy of the emitted photons. Therefore, graphene will be considered as quasi-neutral in the following.

Individual NCs and NPs were excited using a pulsed supercontinuum laser, with a repetition rate tunable from $1.95~\rm MHz$ up to $78~\rm MHz$. The unpolarized output of the supercontinuum laser at a wavelength of $480~\rm nm$ (photon energy of 2.53~eV) was selected using an acousto-optic tunable filter. The full width at half maximum of the filtered pulses was $\approx~50 \rm~ ps$. Wide field PL images were recorded using an electron-multiplying charge coupled device camera (emCCD). PL time traces and PL decays of individual colloidal semiconductor nanostructures were measured in a confocal arrangement,  using an avalanche photodiode coupled to a time-tagged, time-correlated single photon counting board. PL spectra were recorded using a monochromator coupled to a CCD matrix. A pulse fluence lower than $1\times 10^{13} ~ \rm{photons/pulse/cm^{2}}$ at $\lambda=480 ~ \rm{nm}$ was used for all measurements. Considering similar absorption cross sections of a few $10^{-14} ~ \rm{cm^2}$ for our individual NCs and NPs at 480~nm~\cite{leatherdale2002,park2011,she2014}, we can estimate that on average, significantly less than one exciton per incoming laser pulse  is formed in an individual NC or NP. We further verified that the PL decays of individual NCs and NPs are independent on the pulse fluence, in the range $(\approx 10^{12} - 3\times 10^{13} ~ \rm{photons/pulse/cm^{2}})$.

\paragraph{\textbf{Energy transfer on bare graphene}}

Figure~\ref{Fig01} shows wide-field PL images  of individual CdSe/CdS/ZnS NPs deposited on a bare, mechanically exfoliated graphene monolayer.  The PL intensity is strongly quenched, by more than one order of magnitude for NPs deposited on graphene. Very similar results were obtained using CdSe/CdS NCs. As previously discussed for individual core/shell CdSe/ZnS NCs~\cite{chen2010}, we attribute PL quenching to F\"orster-type RET.

We then compare the typical PL spectrum, PL time trace and PL decay of individual NCs (see Figure~\ref{Fig02}) and NPs (see Figure~\ref{Fig03}), measured on a fused quartz substrate and on a bare graphene monolayer. For each nanoemitter investigated, we introduce the average number of emitted photons per incoming laser pulse $N_{\rm em}$ (see PL time traces in Figures ~\ref{Fig02}b,e and \ref{Fig03}b,e). First, we note that although the peak energy of the PL spectra exhibits a slight dispersion over a collection of nanoemitters, we did not observe systematic spectral shifts for NCs or NPs on graphene with respect to a reference on fused quartz. In both cases, the PL count rates on graphene and fused quartz are similar, but the PL signals are obtained using very different repetition rates (compare Figures~\ref{Fig02}b and~\ref{Fig02}e, and Figure~\ref{Fig03}b and in Figure~\ref{Fig03}e, for NCs and NPs, respectively). On these selected examples, $N_{em}$ is quenched by a factor of approximately $50$ $(60)$ when the NC (NP) is adsorbed on graphene, as compared to a reference recorded on fused quartz. Over time scales larger than $100~\rm ms$, we also observe, in agreement with previous observations~\cite{chen2010,ajayi2014} that the blinking behavior, characteristic of NCs and NPs deposited on fused quartz, is seemingly reduced when the nanoemitters are adsorbed on graphene. This observation is presumably due to the acceleration of the excited state decay, which occurs before charge carriers may be trapped and allow the observation of dark and/or grey states~\cite{cichos2007,spinicelli2009,malko2012,tessier2012,tessier2013}.

We now compare the PL decays of individual NCs and NPs measured on fused quartz and on graphene. Examples are shown in Figures ~\ref{Fig02}c and~\ref{Fig02}f for individual NCs and in Figures ~\ref{Fig03}c and~\ref{Fig03}f, for individual NPs, respectively. At this point, let us note that the room temperature typical PL decay of a single NC or NP is not mono-exponential~\cite{spinicelli2009,malko2012,tessier2012,tessier2013}. In the case of NCs, we found that most decays could be well fit to a bi-exponential form, while a stretched exponential form was providing better fits in the case of NPs. These complex behaviors are attributed to the existence of a distribution of bright, grey and dark states with distinct lifetimes~\cite{cichos2007,spinicelli2009,malko2012,tessier2012,tessier2013}. The fractional weight of these states in a measured decay may vary significantly from particle to particle, reflecting heterogeneities in core and shell passivation.  In order to provide a general definition, the measured PL decay rate $\gamma$ is defined as the maximum number of recorded counts divided by the (background corrected) total area of the PL decay. The resulting values are then multiplied by a correction factor, which takes into account the minor contribution of the instrument response function (see Figures ~\ref{Fig02}c,f, Figures \ref{Fig03}c,f, and Supporting Information). We have verified that our conclusions are independent of the method used to define the PL decay rate.
For the individual NC (NP) considered here, we find that $\gamma$ is enhanced by a factor of approximately $50$ ($80$). Remarkably, the quenching factors estimated from the PL time traces are in good agreement with the enhancement factors of the PL decay rate. This suggests that the strong PL quenching is solely due to an increase of the non-radiative decay rate, and that possible modifications of the radiative decay rate of an individual NC or NP adsorbed on graphene can be neglected. This is consistent with recent theoretical calculations, which demonstrated that the radiative lifetime of an individual emitter is marginally modified in the vicinity of graphene~\cite{gomez2011}. Thus, in the following, we will consider that $\gamma\approx\gamma_{t}^{}+\gamma_0$, where $\gamma_t$ is the RET rate and $\gamma_0$ is the reference decay rate measured in the absence of graphene.

Similar measurements were repeated on more than 100 NCs and NPs deposited on bare graphene. We found similar statistically averaged quenching factors of approximately 50 for NCs and NPs. Overall, for $95~\%$ of the investigated single NCs and NPs, the RET efficiency, defined as $\eta=1-\gamma/\gamma_0$ is found to be larger than $95~\%$.

\paragraph{\textbf{Distance dependence of the RET rate}}

\begin{figure*}[hbt]
\begin{center}
\includegraphics[scale=0.68]{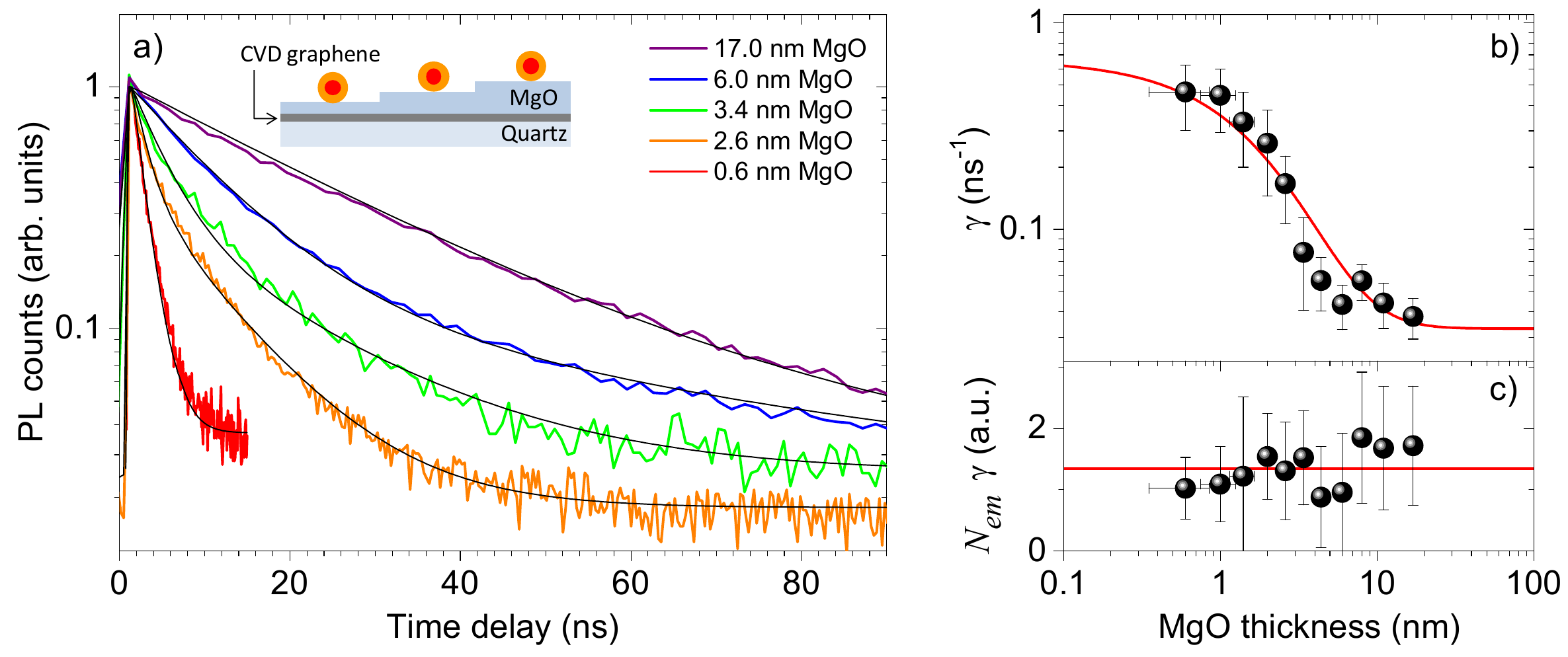}
\caption{a) Selected luminescence decays of individual CdSe/CdS nanocrystals separated from graphene by a MgO spacer graphene with increasing thicknesses. The thin black lines are fits based on bi-exponential decays convoluted with the instrument response function. b) Statistically averaged measured decay rate $\gamma$ as a function of the thicknesses of the MgO spacer. The red solid line is a fit based on Eq.~\ref{eqRET1}, with $p=4$, $d_0=5.5~\rm nm$, $z_0=11.5~\rm nm$. c) Statistically averaged product of the number of emitted photons per exciting laser pulse $N_{\rm em}$ and the decay rate $\gamma$.}
\label{Fig04}
\end{center}
\end{figure*}

\begin{figure*}[hbt]
\begin{center}
\includegraphics[scale=0.68]{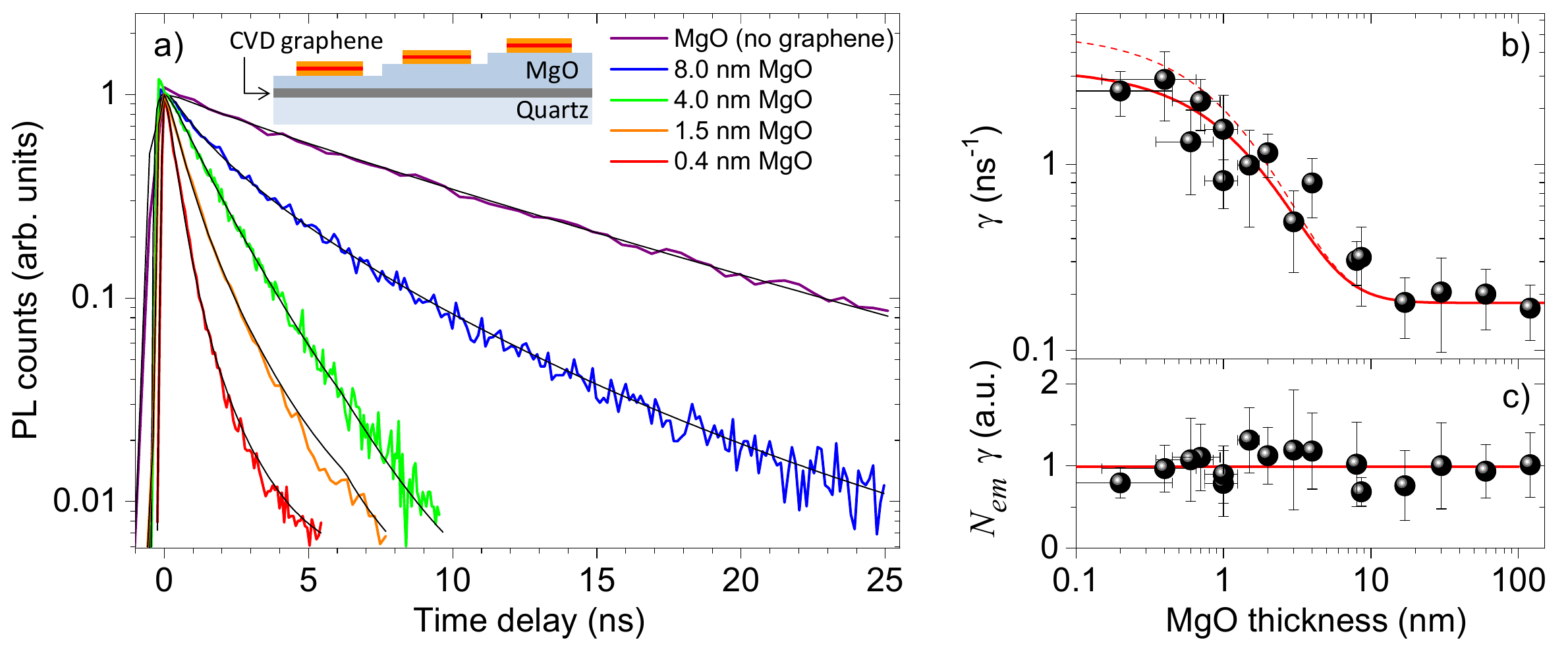}
\caption{a) Selected luminescence decays of individual CdSe/CdS/ZnS nanoplatelets separated from graphene by a MgO spacer graphene with increasing thicknesses. The thin black lines are fits based on stretched exponential decays convoluted with the instrument response function. b) Statistically averaged measured decay rate $\gamma$ as a function of the thicknesses of the MgO spacer. The red solid and dashed lines are fits based on Eq.~\ref{eqRET2D}, with $d_0=3.5~\rm nm$, $\Lambda=7.5~\rm nm$, and on Eq.~\ref{eqRET1} with $p=4$, $d_0=3.5~\rm nm$, $z_0=8.0~\rm nm$, respectively. c) Statistically averaged product of the number of emitted photons per exciting laser pulse $N_{\rm em}$ and the decay rate $\gamma$.}
\label{Fig05}
\end{center}
\end{figure*}

We have measured the PL decay of NCs and NPs separated from a graphene monolayer by a MgO thin film, with a thickness ranging from a few $\angstrom$ up to several tens of nanometers. In these experiments, the reference decay rate $\gamma_0^{}$ is the statistically averaged decay rate of NCs and NPs measured on a bare $>100~\rm nm$ thick film of MgO. The results for NCs and NPs are summarized in Figures \ref{Fig04} and \ref{Fig05}, respectively. Each point corresponds to a statistical average over $10$ to $30$ single NCs, and over more than $25$ single NPs, respectively. The vertical error bars correspond to the standard deviations, while the horizontal error bars account for the roughness of the MgO film. For both NC and NPs, we observe that the measured decay rate decreases significantly, when increasing the thickness of the MgO film. However, as shown in figures \ref{Fig04}c and \ref{Fig05}c, the product $N_{\rm em} \gamma$ varies by less than a factor of 2 and, considering the standard deviations associated with each distribution, can be considered as constant. This generalizes the conclusions drawn from the analysis of Figures~\ref{Fig02} and~\ref{Fig03}.

A key observation is that the decrease of the PL decay rate as function of the MgO thickness is seemingly steeper for NCs than for NPs (see also Figure 7 in the Supporting Information for a comparison of the normalized decay rates). This points towards the effect of dimensionality on RET, which we now discuss. Let us first focus on the case of NCs interacting with graphene. Based on the well known $1/d^{6}$ distance dependence of the F\"orster energy transfer rate between two point-like dipoles~\cite{forster1948}, one would expect the measured decay rate to scale as
\begin{equation}
\label{eqRET1}
\gamma = \gamma_0 \left[ 1+\left( \frac{z_0}{d_0+d_{\rm MgO}^{}} \right)^p \right],
\end{equation}
where the distance $d$ separating the nanoemitter from the graphene layer is $d=d_0+d_{\rm MgO}^{}$, with $d_{\rm MgO}^{}$, the thickness of the MgO film and $d_0$, the minimal distance between the center of the nanoemitter and the graphene surface in the absence of a MgO spacer, where $z_0$ characterizes the RET efficiency, and $p$ is related to the dimensionality of the donor and acceptor. A straightforward extension of F\"orster's theory would then give $p=4$ for a zero-dimensional emitter interacting with a two-dimensional assembly of independent dipoles~\cite{kuhn1970,chance1978}. Indeed, for NCs (see Fig.~\ref{Fig04}b, using $p=4$, we obtain a good fit with $z_0=(11.5\pm~1.5)\rm~nm$ and $d_0=(5.5\pm~1)\rm~  nm$. The latter value is slightly larger than the average physical radius of the NCs, which is consistent with a possible contribution from the surrounding ligands and residual adsorbates to $d_0$. 

Although a fit based on Eq.\ref{eqRET1} is in good agreement with our measurements, one has to recall that Eq.\ref{eqRET1} overlooks the fact that graphene is a two-dimensional system with extended, delocalized electronic wavefunctions and a well defined electronic dispersion. The RET rate $\gamma_{t}^{0\rm D}$ from a single point-like dipole to  graphene has been extensively studied theoretically in recent years~\cite{swathi2008,swathi2009,gomez2011,velizhanin2011,gaudreau2013,malic2014} and can be written as~\cite{swathi2009,gomez2011}

\begin{equation}
\label{eqRET0D}
\gamma_{t}^{0 \rm D} \propto \int_{0}^{q_{\rm max}^{}}{\frac{\mathrm{d}q\; q^3 e^{-2qd}}{\sqrt{q_{\rm max}^{2}-q^2}}},
\end{equation}
where $q$ is the transferred momentum, $q_{\rm max}^{}=\frac{2\pi c}{\lambda_0 v_F^{}}$ is the largest transferable momentum, $\lambda_0$ is the wavelength of the emitted photons, $c$ is the speed of light and $v_F^{}\approx1.1\times 10^6 ~\rm m/s$ is the Fermi velocity in graphene. Contrary to far-field emission and absorption processes, the RET process described by Eq.\ref{eqRET0D} involves a finite momentum transfer (see Figure~\ref{Fig01}a), which is on the order of $1/d$. For neutral graphene, and distances such that $d\gg1/q_{\rm max}^{}\approx0.3~\rm nm$, the denominator of the integrand in Eq.\ref{eqRET0D}, which originates from the (momentum dependent) optical conductivity of graphene~\cite{gomez2011}, only gives significant contributions for $q\ll q_{\rm max}^{}$ (quasi-vertical transitions), and can thus be approximated as a constant. In these conditions, graphene can be treated as a two-dimensional assembly of incoherent point-like dipoles, and Eq.\ref{eqRET0D}~simplifies as $\gamma_{t}^{0\rm D}\approx \gamma_0 \left(\frac{z_0^{}}{d}\right)^4$. Nevertheless, it must be emphasized that Eq.~\ref{eqRET0D} applies to a point-like donor, while in our case, the NCs have a finite radius, larger than the thinnest MgO films ($0.6~\rm nm$) deposited here. However, the surrounding ligands and the finite thickness of the CdS shell of our CdSe/CdS NCs warrant that there is a minimal distance of a few nanometers between the graphene layer and the emitting CdSe core. We will therefore consider that the \textit{long distance} approximation is valid and neglect the finite size of the NCs and NPs. Considering random relative dipole orientation, we obtain~\cite{swathi2009} $z_0^4 = \frac{3 \alpha}{2048\;\pi^3 F^2 \varepsilon^{5/2}}  \lambda_0^{4}$, where $\alpha\approx1/137$ is the fine structure constant, $\varepsilon$ is the effective dielectric constant (at $\lambda_0$) of the surrounding medium and $F=\frac{3\varepsilon}{\varepsilon_{\rm CdS}^{}+2\varepsilon}$ is a screening factor~\cite{califano2005}. Using $\varepsilon\approx (\varepsilon_{\rm \rm MgO}^{}+1)/2\approx 2$ results in $z_0\approx 11~\rm nm$, in good agreement with our measurements.

We now address the distance dependence of the RET rate from single NPs to graphene. The NPs cores are atomically smooth and have recently been shown to behave as genuine quantum wells~\cite{ithurria2011}. As a result, a different regime of RET is anticipated when a two-dimensional donor is involved. Electronic states in NPs are described by extended wave functions that are coherent over large distances in the NP plane. However, at finite temperature, one has to consider, in a photoexcited NP, a thermal distribution of excitons with different in-plane center of mass momentum. Besides, since the thickness of the NP shell is on the order of $2~\rm nm$, there is a minimal separation between the NP core and graphene, such that the approximation $d\gg1/q_{\rm max}$ holds (\textit{i.e.}, graphene can also be treated as an incoherent plane of point-like-dipoles). Similar situations have been previously modeled by Basko \textit{et al.}~\cite{basko2000} and Kos \textit{et al.}~\cite{kos2005} for hybrid systems composed of quantum wells transferring energy to an assembly (thick film or monolayer) of point-like dipoles. The calculated RET rate writes
\begin{equation}
\gamma_{t}^{2\rm D}(d)\propto \int_{0}^{\infty}{\mathrm{d}q\; q^3 e^{-2qd}e^{-\left(\frac{\Lambda q}{2\pi}\right)^2}},
\label{eqRET2D}
\end{equation}
where $\Lambda= \frac{h}{\sqrt{2m^{\ast}k_{\rm B}^{}T}}$ is the de Broglie thermal length, with $h$ the Planck constant, $k_{\rm B}$ the Boltzmann constant and $m^{\ast}$ the mass of the lowest energy heavy hole exciton ($m^{\ast}\approx m_{\rm e}$, where $m_{\rm e}$ is the free electron mass) in CdSe~\cite{ithurria2011,LandoltBornstein}.

This expression applies to free excitons in an infinite quantum well, with thickness $(L_z\ll d)$. The assumption of free excitons is consistent with the reported low density of trapping sites of CdSe-based core/shell NPs~\cite{tessier2013}. In addition, since the typical lateral dimensions of our NPs ($L_x= (22~\pm 2) \rm~ nm$ and $L_y = (9~\pm 1.5) \rm~ nm$) exceed $\Lambda$ (at room temperature, $\Lambda\approx 7.5~\rm nm$), finite size effects can be neglected. We have thus attempted to fit the data in Figure~\ref{Fig05}b using the expression $\gamma\left(d\right)=\gamma_0+\gamma_{t}^{2 \rm D}\left(d\right)$. A good agreement with our experimental results is obtained using $d_0=\left(3.5\pm 1\right) ~\rm nm$. Again, the latter value is consistent with the thickness of the core and shell of the NPs and includes a minor contribution from the surrounding ligands and residual adsorbates.

An analysis of the limiting cases of  Eq.~\ref{eqRET2D} provides a rationale  for the observed scaling. In the short distance limit, $d\ll \Lambda$, $\gamma_{t}^{2D}$ becomes independent on the distance and is determined by the thermal cutoff. In contrast, in the large distance (or high temperature) limit, $d\gg \Lambda$ and $\gamma_{t}^{2 \rm D}$ follows a $1/d^4$ scaling, as expected in the case of a two-dimensional assembly of incoherent point-like dipoles. Since, at room temperature, $\Lambda$ falls exactly in our measurement range, we expect the RET rate to decay more smoothly than $1/d^4$ for $d<\Lambda$. This is indeed observed experimentally (see Fig~\ref{Fig05}b), since a scaling based on Eq.~\ref{eqRET0D}, with $p=4$ and $d_0=3.5~\rm nm$, having asymptotic behavior for large $d$ as our fit based on Eq.~\ref{eqRET2D}, would predict a higher $\gamma$ in the short distance limit. 

The results shown in Figure~\ref{Fig05} suggest that even at room temperature, the distance scaling of the RET rate from an extended donor (such as a two-dimensional NP) to graphene exhibits, as expected theoretically, a slight deviation from a simple power law. Following Eq.\ref{eqRET2D}, this deviation is expected to be more prominent at lower temperatures. Thus, further investigations of the distance and temperature dependence of the RET rate for various NP architectures could provide insights into the dimensionality of excitons in these novel systems~\cite{rindermann2011}.

In conclusion, F\"orster-type resonant energy transfer dramatically affects the photophysics of semiconductor nanostructures absorbed on graphene. In spite of its atomic thinness, a single layer of graphene typically quenches more than $95\%$ of the luminescence of individual nanocrystals or nanoplatelets. The observation of well-defined distance scalings of the resonant energy transfer rate suggests that novel graphene-based molecular rulers can be engineered using semiconductor nanostructures with different size, shape, and dimensionality. This is a promising development, especially for biological research. Finally, with the prospect of designing hybrid opto-electronic devices, we show that graphene can very efficiently harvest energy from photoexcited semiconductor nanostructures, which is of interest for photodetection. A major challenge is now to dissociate the electron-hole excitations generated in graphene before their fast relaxation into heat~\cite{johannsen2013,gierz2013}.

\paragraph{\textbf{Acknowledgement}} 
We are grateful to D.M. Basko and G. Weick for fruitful discussions, to R. Bernard, S. Siegwald and H. Majjad for help with sample fabrication and characterization in the STNano clean room facility, and to M. Romeo for technical support. The authors at IPCMS acknowledge financial support from the CNRS, Universit\'e de Strasbourg, C'Nano GE and the Agence Nationale de Recherche (ANR) under grants QuandDoGra ANR-12-JS10-0001 and  Fungraph ANR-11-IS10-0003. B.D. (at ESPCI) thanks the ANR for funding under grants SNAP and QDOTICS. J-O. L. (at KRICT) acknowledges support from NRF-ANR program through the National Rearch Foundation of Korea funded by the Ministry of education, Science and Technology (NRF-2011-K2A1A5-2011-0031552). 

\paragraph{\textbf{Supporting Information}} 
Synthesis and characterization of CdSe/CdS nanocrystals and CdSe/CdS/ZnS nanoplatelets, graphene growth by chemical vapor deposition, sample characterization by micro-Raman spectroscopy, definition of the energy transfer rate, comparison of the normalized distance dependent decay rates.


%


\onecolumngrid
\newpage
\begin{center}
{\Large\textbf{Supporting Information}}
\end{center}

\setcounter{equation}{0}
\setcounter{figure}{0}
\setcounter{section}{0}
\renewcommand{\theequation}{S\arabic{equation}}
\renewcommand{\thefigure}{S\arabic{figure}}
\renewcommand{\thesection}{S\arabic{section}}


\makeatletter
\DeclareRobustCommand{\element}[1]{\@element#1\@nil}
\def\@element#1#2\@nil{%
  #1%
  \if\relax#2\relax\else\MakeLowercase{#2}\fi}
\pdfstringdefDisableCommands{\let\element\@firstofone}
\makeatother

\section{Synthesis of \element{Cd}\element{Se}/ \element{Cd}S nanocrystals}
\paragraph{\textbf{Chemicals:}} 1-Octadecene (ODE, $90~\%$, Aldrich), oleylamine (OLA, $70~\%$, Fluka), oleic acid (OA, $90~\%$, Aldrich), sodium myristate ($99~\%$, Fluka), cadmium oxide ($99.99~\%$, Aldrich), selenium powder 100 mesh ($99.99~\%$, Aldrich), sulfur ($99.998~\%$, Aldrich), trioctylphosphine (TOP, $90~\%$, Cytec) and tetradecylphosphonic acid (TDPA, $97~\%$, PCI synthesis) were used as received.

\paragraph{\textbf{Precursors preparation:}}
Cadmium myristate was prepared according to ref.~\cite{yang2005}. The solution of cadmium oleate 0.5 M in oleic acid was synthesized by heating $6.42$~g of CdO in 100 mL of oleic acid at $160~^{\circ}\rm C$  under argon until it turned colorless. The solution was then degassed under vacuum at $100~^{\circ}\rm C$ for 1 hour. Sulfur stock solution in ODE (S-ODE $0.1$~M) was prepared by heating 480~mg of sulfur in 150~mL of degassed ODE at $120~^{\circ}\rm C$ until complete dissolution.  TOP-Se 1 M in TOP was prepared by dissolving $15.8$~g of Se powder in 200~mL TOP under magnetic stirring overnight in a glove box. 

\paragraph{\textbf{Synthesis of CdSe cores:}} CdSe nanocrystals (NCs) were prepared by a procedure slightly modified adapted by Mahler \textit{et al.}~\cite{mahler2008} from Mohamed \textit{et al.}~\cite{mohamed2005}. A mixture of 2~mL of Cd(oleate)$_2$ (0.5~M) and 3~mL of ODE was degassed under vacuum at $70~^{\circ}\rm C$ during 30~min and heated under argon flow up to $240~^{\circ}\rm C$. A mixture of $1.5$~mL of TOP-Se 1~M, $1.5$~mL of oleylamine and 1~g of TDPA was heated until complete dissolution then injected and the solution was annealed for 30 seconds at $190~^{\circ}\rm C$, and then immediately cooled down to room temperature. The solution was washed up with 40~mL of ethanol. The solution was centrifuged at 5500~RPM in order to precipitate the TDPA. The nanocrystals were suspended in 20~mL of toluene, washed again in 20~mL of ethanol, and dispersed in 10~mL of hexane. The nanocrystals obtained with this protocol were around 2 nm diameter, and their approximate concentration was $80~\mu\rm M$.

\paragraph{\textbf{Synthesis of CdS shell on the CdSe cores:}} For the CdS shell growth on the CdSe cores, a mixture of 3.1 mL of solution of cores dispersed in hexane, 5 mL ODE and 50 mg Cd(myr)2 was degassed under vacuum, at $70~^{\circ}\rm C$ for 30 minutes, and then put under argon flow. The temperature set value was then increased to $300~^{\circ}\rm C$, and when the temperature reached $100~^{\circ}\rm C$, 1~mL of OLA was injected, followed by a mixture of 8~mL SODE (0.1~M), 1.6 mL Cd(OA)$_2$ (0.5~M) and 1~mL OLA at an injection rate of first 2~mL/h for 2~mL, then 16~mL/h for the rest of the seringe. Once the injection was finished, a mixture of 0.5~mL OLA and 0.5 mL Cd(OA)$_2$ 0.1~M diluted in OA was added, and the solution was annealed for 10 minutes at $300~^{\circ}\rm C$. The solution was then cooled down to room temperature and the nanocrystals were washed with ethanol, centrifuged and redispersed in 10~mL hexane. 

\begin{figure*}[hbt]
\begin{center}
\includegraphics[scale=0.62]{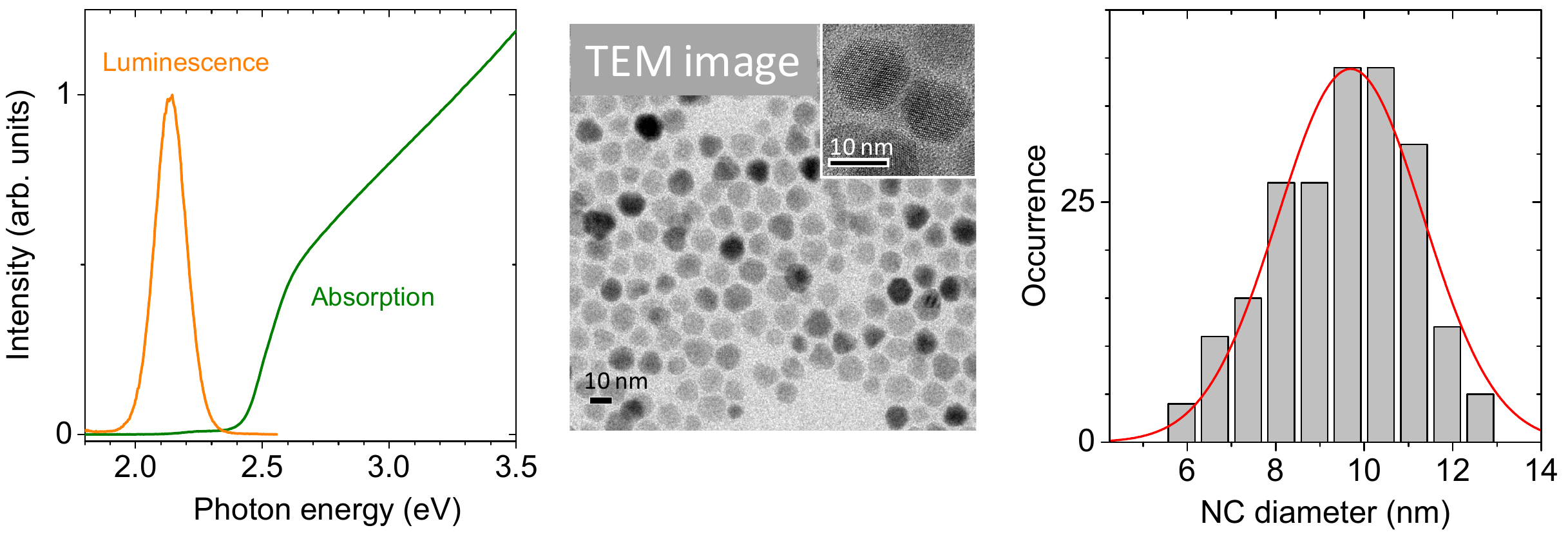}
\caption{(left) Ensemble absorption and photoluminescence spectra of the CdSe/CdS nanocrystals used in our measurements. (center) Transmission electron microscope image of the CdSe/CdS nanocrystals. (right) Histogram of the NC diameters.}
\label{FigS1}
\end{center}
\end{figure*}

\paragraph{\textbf{NC characterization:}} The ensemble absorption spectrum was measured on a Cary 5E UV-visible spectrometer. The ensemble photoluminescence (PL) spectrum was acquired with an Edinburgh Instruments FCS900 spectrometer. Transmission electron microscopy (TEM) images were taken using a TEM JEOL 2010 with field emission electron gun. 
The results for CdSe/CdS NCs are shown in Figure~\ref{FigS1}. The NCs show strong PL at $2.14~\rm eV$, with a full width at half maximum (FWHM) of $140~\rm meV$. From the analysis of the TEM images, we find an statistically averaged diameter of $(9.5\pm 1.5)~\rm nm$.


\section{Synthesis of \element{Cd}\element{Se}/\element{Cd}\element{Se}S/\element{Zn}S nanoplatelets}


\paragraph{\textbf{Preparation of CdSe nanoplatelets}}
CdSe nanoplatelets (NPs) with an atomically controlled thickness of 4 monolayers were prepared as described in ref~\cite{ithurria2008}. In a 100 mL three neck flask 170 mg (0.3 mmol) of cadmium myristate were introduced along with 12 mg (0.15 mmol) of Se powder and 15 mL of octadecene. The solution was degassed under vacuum for 30 min at room temperature. Under Ar flow, the temperature was set at $240~^{\circ}\rm C$. When the temperature reached $200~^{\circ}\rm C$ (the solution is yellow orange at this step), 40 mg (0.15 mmol) of cadmium acetate were quickly added to the solution. Finally the reaction was left for 12 min at $240~^{\circ}\rm C$. Oleic acid (2 mL) were then injected to quench the reaction, and the flask was cooled down to room temperature. The nanoplatelets were precipitated by adding 15 mL of hexane and 15 mL of EtOH. After centrifugation at 6000 rpm for 10 min, the clear supernatant was discarded, and the solid precipitate was re-dissolved in fresh hexane (8 mL). The cleaning procedure was repeated a second time.

\paragraph{\textbf{Preparation of CdSe/(CdS)$\mathbf{_3}$/(ZnS)$\mathbf{_2}$ nanoplatelets}}
CdSe NPs (500 $\mu L$ of the solution previously obtained were charged in a 3 mL vial with 0.5 mL of N-methylformamide (NMF) obtaining a biphasic mixture. Then 20 $\mu L$ from a freshly solution of Na$_2$S 0.3 M in NMF were added to the biphasic system stirring at room temperature for few minutes. The NPs were thus transferred in the polar NMF bottom phase that turned into orange. The hexane phase was discarded and NPs in NMF were washed twice with hexane to remove residual organic ligand. Then, a mixture of toluene: acetonitrile in a ratio 3:1 is added and NPs were precipitated using centrifugation. The NPs were re-dissolved in 0.5 mL of NMF and 30 $\mu L$ of Cd(OAc)$_2$ 0.1 M in NMF were added to complete the first CdS monolayer shell deposition. After stirring of few minutes a room temperature, the NPs were precipitated as described above and re-dissolved in 0.5 mL of NMF. The procedure was repeated two more times for a total of 3 CdS layer deposition. To deposit the two final ZnS shell layers, 20 $\mu L$ of of Na$_2$S 0.3 M in NMF were added to NPs in NMF and NPs were precipitated using toluene:acetonitrile in a ratio 3:1. After redispersion on NPs in 0.5 mL of NMF, 30 $\mu L$ of Zn(OAc)$_2$ 0.1 M in NMF were added and the mixture stirred for few minutes at room temperature. Then, it was precipitated as described above and the procedure was repeated once to complete the shell. The final NPs core/shell were precipitated with toluene and redispersed in 1 mL of hexane adding 100 $\mu L$ of oleic acid and 50 $\mu L$ of oleylamine. The excess of organic ligands was washed away by precipitation with EtOH and finally NPs were dissolved in hexane or toluene to be analyzed by transmission electron microscopy (TEM).  

\begin{figure*}[hbt]
\begin{center}
\includegraphics[scale=0.62]{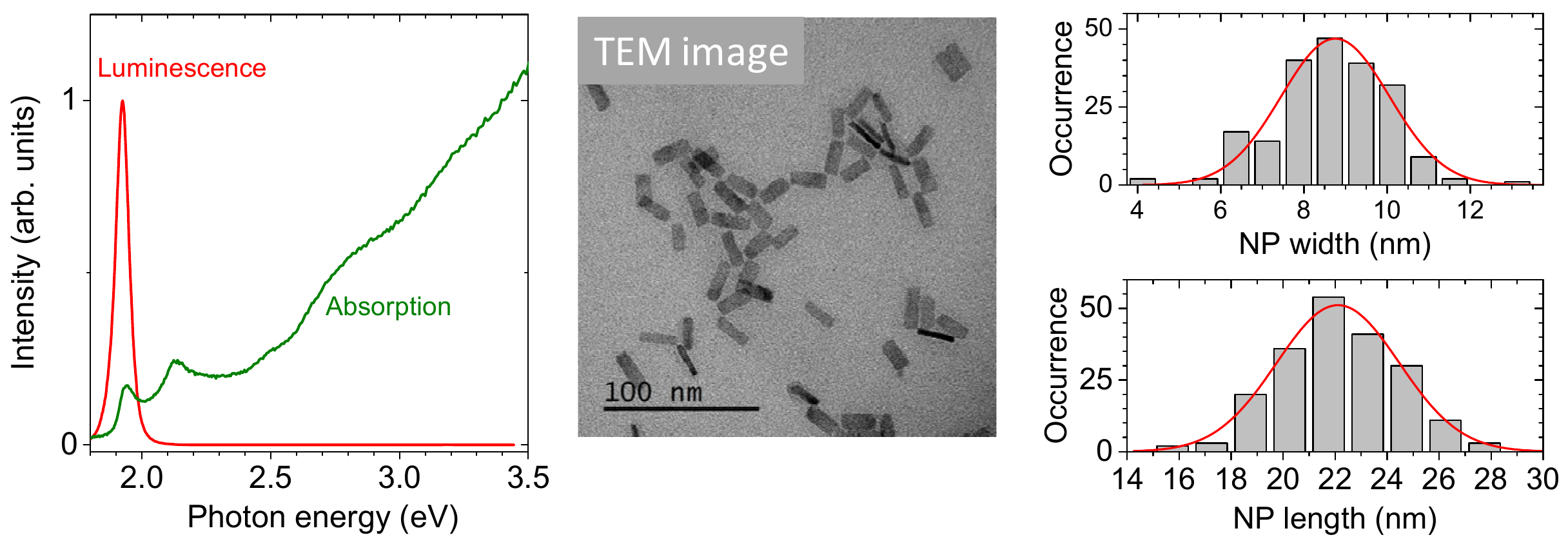}
\caption{(left) Ensemble absorption and photoluminescence spectra of the CdSe/CdS/ZnS nanoplatelets used in our measurements. (center) Transmission electron microscope image of the CdSe/CdS/ZnS nanoplatelets. (right) Histogram of the NP widths and lengths.}
\label{FigS2}
\end{center}
\end{figure*}

\paragraph{\textbf{NP characterization:}}  Optical and TEM characterizations of CdSe/CdS/ZnS NPs used in our measurements were performed using the same instruments as for the CdSe/CdS nanocrystals. Figure~\ref{FigS2} shows the absorption and PL spectra, as well as a typical TEM image. These NPs show strong emission at $1.925~\rm eV$, with a FWHM of $60~\rm meV$. From the analysis of the TEM images, we find an average lateral dimensions of $(9\pm 1.5)~\rm nm \times (22\pm 2)~\rm nm$. The average height of the NPs is estimated to be $(3.5\pm 0.5)~\rm nm$.


\section{Graphene grown by chemical vapor deposition}
	To investigate the distance scaling of the energy transfer rate we made use of graphene grown by low-pressure chemical vapor deposition (LPCVD) on a Cu foil (Alfa Aesar, Item No. 46986, $99.8~\%$, cut into 6 x 6 cm strips) in a hot wall furnace consisting of a 4 inch fused silica tube~\cite{li2009}. Prior to CVD, the foils were cleaned using a Ni etchant for 5 min and then thoroughly rinsed with DI water. A typical growth process flow is as follows: (1) load the Cu foil, evacuate, heat to $1000~^{\circ}\rm C$,~and anneal for 20 min under a 100 sccm H$_2$ flow (at a pressure of about $70-80~\rm mTorr$); (2) introduce 30 sccm CH$_4$ and 30 sccm H$_2$ for 40 min($60~\rm mTorr$); (3) final exposure to CH$_4$ for 40 min, followed by a cool down of furnace to room temperature in vacuum. A poly(methyl methacrylate) (PMMA) solution (950 K, $4\%$ by volume dissolved in chlorobenzene) was spin-coated onto the top side of the sample followed by baking at $60~^{\circ}\rm C$ for 5 min. The Cu under the graphene film was etched using a copper etchant solution and washed with DI water 3 times. The resulting PMMA/grapheme film is transferred onto a fused quartz substrate and the PMMA film is dissolved using acetone. In order to remove resist residues and other chemical contaminants from the graphene surface, the samples were heated at $250~^{\circ}\rm C$  for 4 hours in a tube furnace in a Ar/H$_2$ ($90/10~\%$ mixture composition) atmosphere.

\section{Sample characterization by Raman spectroscopy}

\begin{figure*}[!htb]
\begin{center}
\includegraphics[scale=0.68]{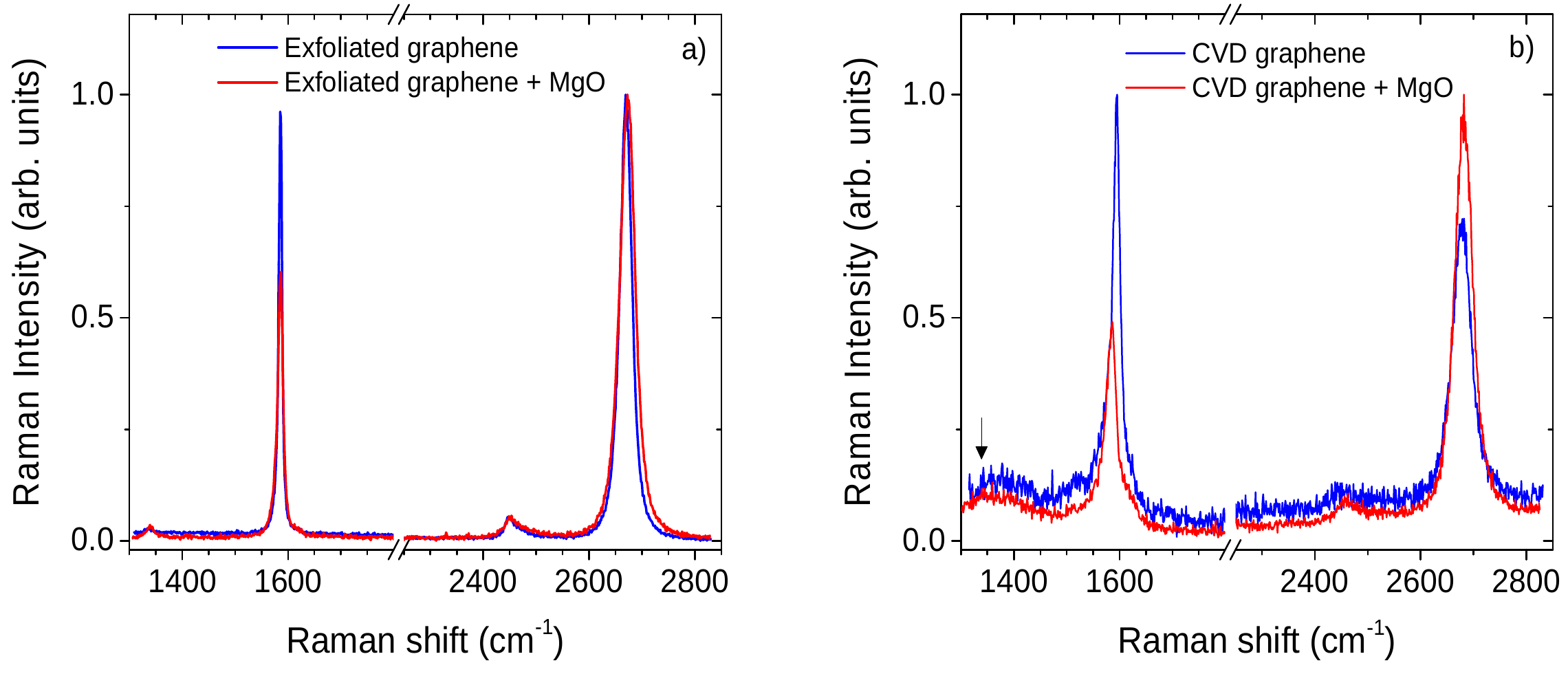}
\caption{Raman spectra of (a) exfoliated graphene and (b) graphene grown by chemical vapor deposition before and after deposition of a thin MgO film. The arrow in b) indicates the expected position of the D mode feature, which is presumably drown into the Raman background in these measurements.}
\label{FigS3}
\end{center}
\end{figure*}

Micro-Raman measurements were performed using a home-built setup, with a laser photon energy of 2.33 eV (532 nm) and a power of a few hundred $\mu$W focused onto a diffraction limited spot of $\sim 0.6~\rm \mu m$ diameter.  Figure~\ref{FigS3}a shows Raman spectra of a mechanically exfoliated graphene monolayer before and after deposition of a thin MgO layer by means of molecular beam epitaxy~\cite{godel2013}. The narrow and quasi symmetric lineshape of the 2D mode feature (frequency $\omega_{2D}^{}=2670~\rm cm^{-1}$, FWHM $\Gamma_{2D}^{}=28~\rm cm^{-1}$) is a fingerprint of a graphene monolayer~\cite{ferrari2013}.  The bare graphene monolayer exhibits a G-mode frequency (FWHM) of $1586~\rm cm^{-1}$ ($7~\rm cm^{-1}$) that are indicative of a slight unintentional doping, on the order of $10^12 ~\rm cm^{-2}$. This translates into a shift of the Fermi level of less than $200~\rm meV$ relative to the Dirac point~\cite{das2008}.

\begin{figure*}[!htb]
\begin{center}
\includegraphics[scale=0.75]{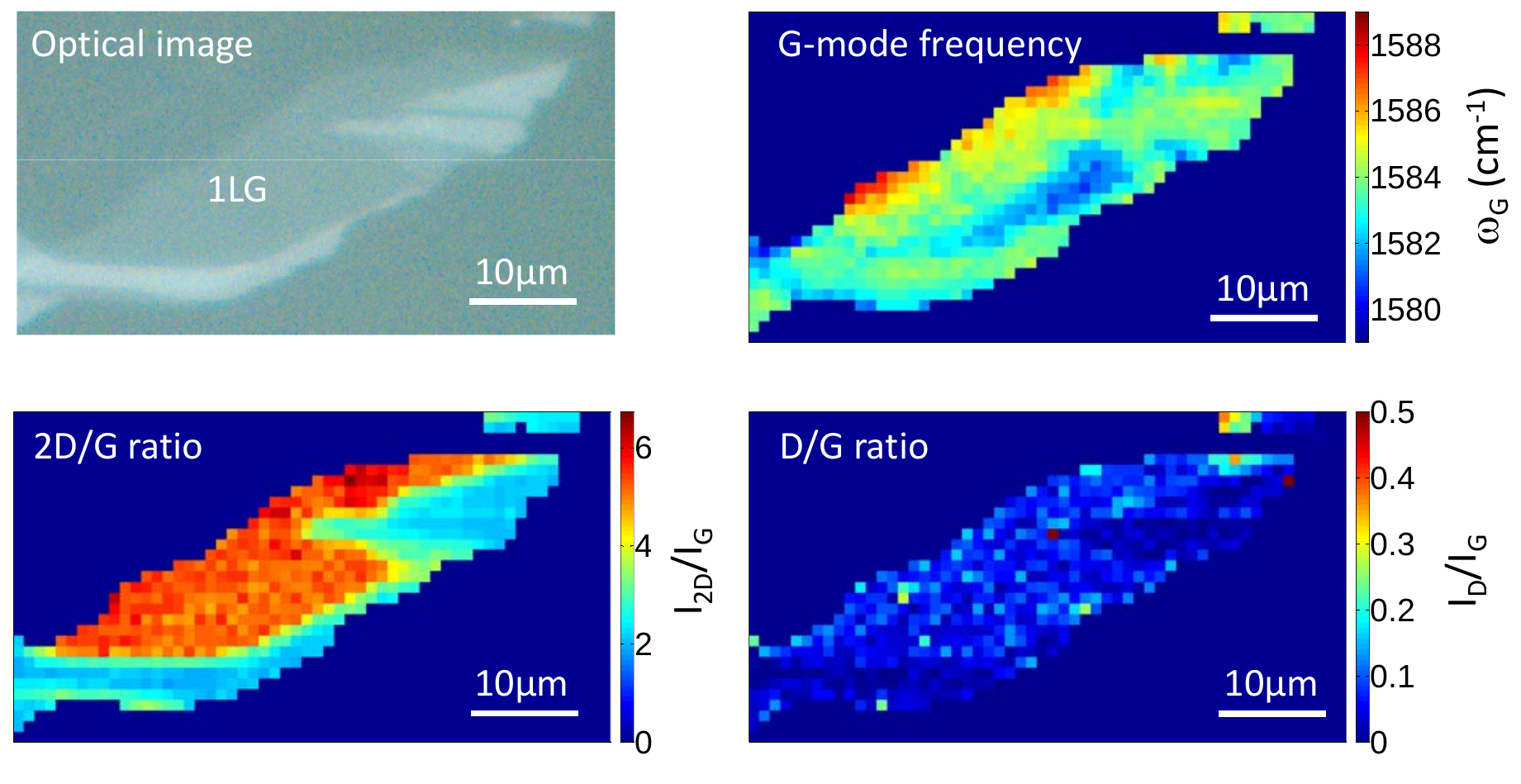}
\caption{Optical image and spatially resolved Raman study of a mechanically exfoliated  graphene sample after deposition of a 2.2 nm thick MgO film.}
\label{FigS4}
\end{center}
\end{figure*}

\begin{figure*}[!htb]
\begin{center}
\includegraphics[scale=0.75]{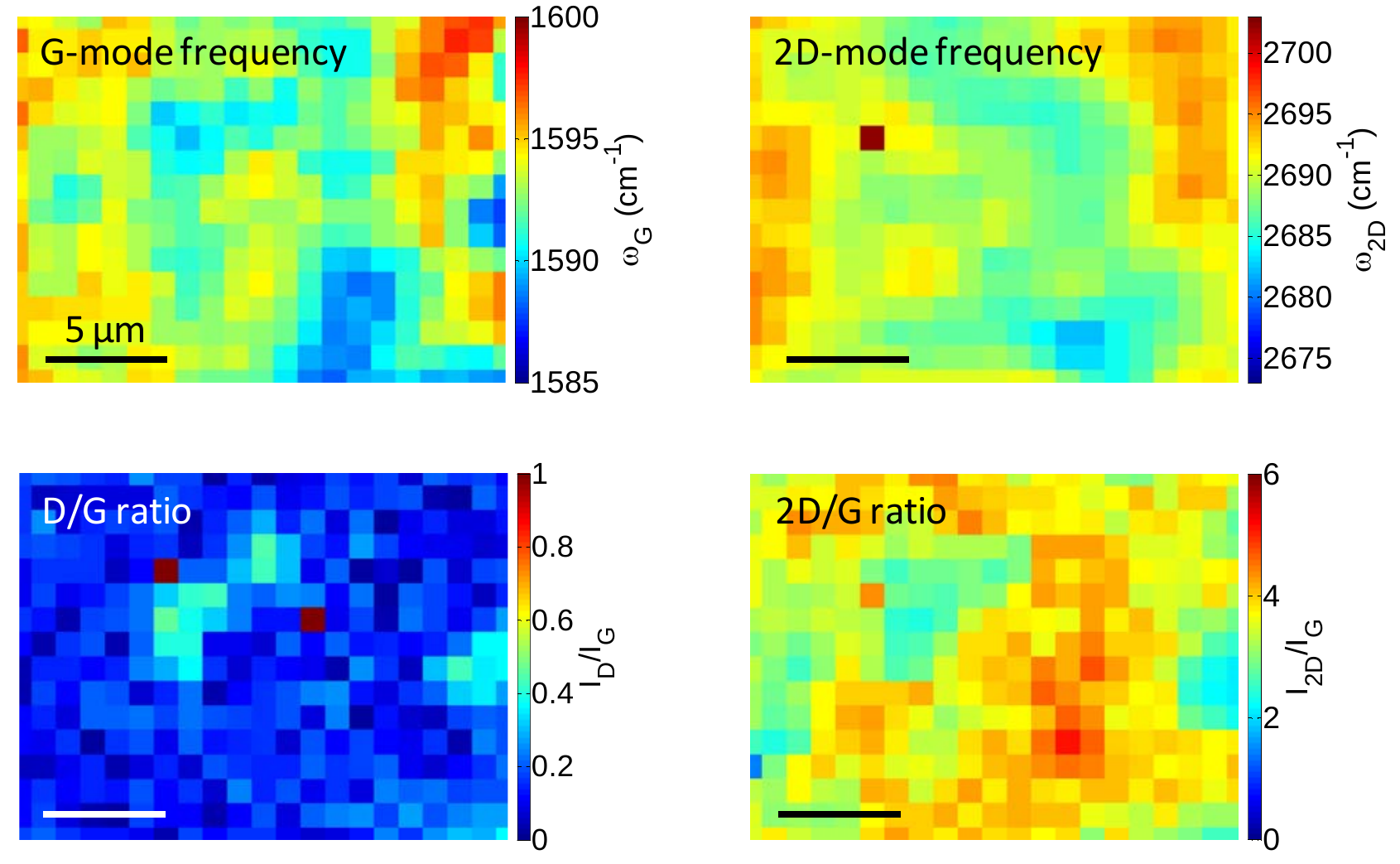}
\caption{Spatially resolved Raman study of a CVD graphene sample after deposition of a 3.2 nm thick MgO film.}
\label{FigS5}
\end{center}
\end{figure*}

After deposition of MgO, we find very similar G and 2D mode frequencies and a slight broadening of the Raman features. The integrated intensity ratio of the defect-induced D mode and the G mode features increases moderately from $I_{\rm D}^{}/I_{\rm G}^{}\sim 1\%$ up to $I_{\rm D}^{}/I_{\rm G}^{}\sim 3\%$, but remains low (see also Figure~\ref{FigS4}. We conclude that the deposition of MgO has no major impact on the doping level and is not introducing significant strain as evidenced by the very slight changes in the frequencies. Similar conclusions are reached for a CVD graphene sample transferred on fused silica (see Figure~\ref{FigS3}b and Figure~\ref{FigS5}). We observe a slightly stronger background than for measurements on mechanically exfoliated graphene, presumably arising from PMMA residues. CVD graphene also exhibits broader Raman features, with slightly more scattered frequencies (compare Figure~\ref{FigS4} and Figure~\ref{FigS5}) than for mechanically exfoliated graphene. This likely arises from increased disorder and residual charge inhomogeneity in CVD graphene. Nevertheless, the Raman features are not significantly affected by the deposition of MgO. These results justify the suitability of CVD graphene for our measurements.

\newpage

\section{Determination of the decay rates}

\begin{figure*}[!hbt]
\begin{center}
\includegraphics[scale=0.45]{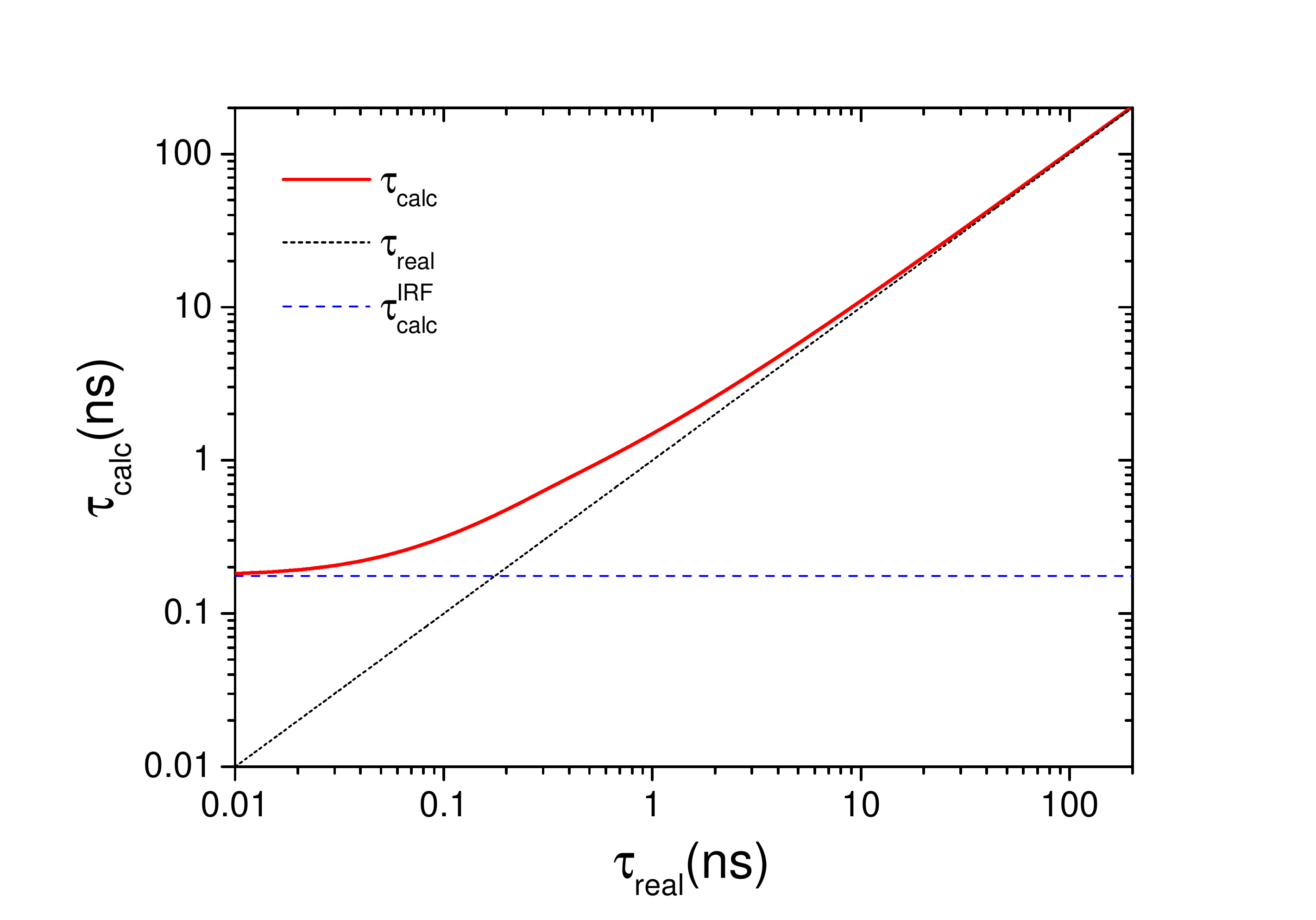}
\caption{Decay time $\tau_{\rm calc}^{}$ numerically computed from the convolution of an exponential decay (decay time $\tau_{\rm real}$) and our instrument response function as a function of $\tau_{\rm real}$.}
\label{FigS6}
\end{center}
\end{figure*}

We define the \textit{calculated} luminescence decay time $\tau_{\rm calc}^{}$ as the ratio between the area under the background corrected PL decay curve and its peak value.
Applying the latter procedure to our instrument response function (IRF) (see Figures 2 and 3 in the main manuscript) yields a value of $\tau_{\rm calc}^{IRF}=176~\rm ps$.
Since the shortest $\tau_{\rm calc}^{}$  values measured for NPs on graphene are below 1~ns, the contribution of the IRF to the PL decay has to be taken into account.
For this purpose, we have computed $\tau_{\rm calc}^{}$ for a set of mono-exponential decays (with a decay time $\tau_{\rm real}$) that have been convoluted with the IRF. The resulting $\tau_{\rm calc}^{}$ are plotted against $\tau_{\rm real}^{}$  in Figure~\ref{FigS6}). 

Then, for each measured PL decay, we calculate $\tau_{\rm calc}^{}$ and estimate a value of $\tau_{\rm real}^{}$, by interpolating the calibration curve shown in Figure~\ref{FigS6}.
Obviously, the obtained correction factor assumes a mono-exponential decay. In practice, since  $\tau_{\rm calc}^{}$ is always significantly greater than $\tau_{\rm calc}^{IRF}$, similar correction factors are obtained using other functional forms (bi-exponential decays or stretched exponential decays). We therefore chose to use the procedure described above in order to obtain a general definition of $\tau_{\rm real}^{}$.

In Figures 4 and 5 of the main manuscript, for each thickness of the MgO spacer, we have defined the average decay rate $\gamma$ as the inverse of the average decay time $\tau_{\rm real}^{}$. We have verified that our conclusions are independent of the method used to define the average PL decay rate.

\newpage

\section{Comparison of the distance dependent decay rates}

\begin{figure*}[!hbt]
\begin{center}
\includegraphics[scale=0.35]{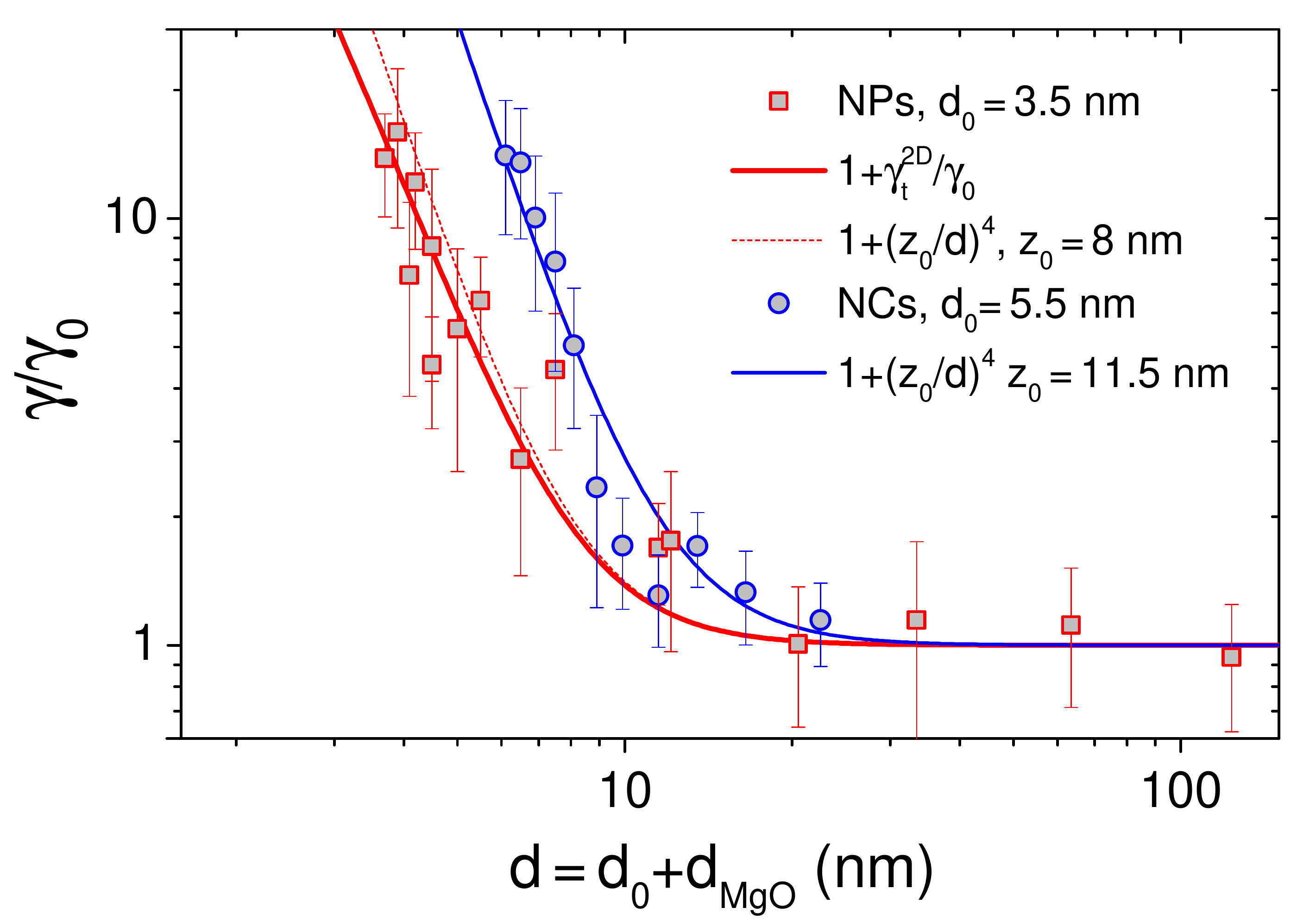}
\caption{Normalized decay rates measured on individual CdSe/CdS NCs (circles) and CdSe/CdS/ZnS NPs (squares) as a function of the total distance $d=d_0^{}+d^{}_{\rm MgO}$ between the nanoemitters and graphene. The solid blue line is computed using Equation 1 in the main text, with $p=4$ and $z_0^{}=11.5~\rm nm$. The solid red line is computed using Equation 3 of the main text, with $\Lambda=7.5~\rm nm$ and a proportionality factor that best fits our results. The dashed red line is computed using Equation 1 in the main text, with $p=4$ and $z_0^{}=8.0~\rm nm$.}
\label{FigS7}
\end{center}
\end{figure*}

Figure~\ref{FigS7} compares the normalized distance decay rates $\gamma/\gamma_0^{}$ measured for NCs and NPs (see Figures 4 and 5 of the main manuscript) as a function of the \textit{total} distance $d=d_0^{}+d_{}^{\rm MgO}$ between the nanoemitters and graphene. Since the minimal distance $d_0^{}$ (see main text) is smaller for NPs than for NCs, our results suggest that at a given \textit{total} distance, the RET rate is larger in the case of a zero dimensional NC than for a two-dimensional NP.



\begin{thebibliography}{62}%
\makeatletter
\providecommand \@ifxundefined [1]{%
 \@ifx{#1\undefined}
}%
\providecommand \@ifnum [1]{%
 \ifnum #1\expandafter \@firstoftwo
 \else \expandafter \@secondoftwo
 \fi
}%
\providecommand \@ifx [1]{%
 \ifx #1\expandafter \@firstoftwo
 \else \expandafter \@secondoftwo
 \fi
}%
\providecommand \natexlab [1]{#1}%
\providecommand \enquote  [1]{``#1''}%
\providecommand \bibnamefont  [1]{#1}%
\providecommand \bibfnamefont [1]{#1}%
\providecommand \citenamefont [1]{#1}%
\providecommand \href@noop [0]{\@secondoftwo}%
\providecommand \href [0]{\begingroup \@sanitize@url \@href}%
\providecommand \@href[1]{\@@startlink{#1}\@@href}%
\providecommand \@@href[1]{\endgroup#1\@@endlink}%
\providecommand \@sanitize@url [0]{\catcode `\\12\catcode `\$12\catcode
  `\&12\catcode `\#12\catcode `\^12\catcode `\_12\catcode `\%12\relax}%
\providecommand \@@startlink[1]{}%
\providecommand \@@endlink[0]{}%
\providecommand \url  [0]{\begingroup\@sanitize@url \@url }%
\providecommand \@url [1]{\endgroup\@href {#1}{\urlprefix }}%
\providecommand \urlprefix  [0]{URL }%
\providecommand \Eprint [0]{\href }%
\providecommand \doibase [0]{http://dx.doi.org/}%
\providecommand \selectlanguage [0]{\@gobble}%
\providecommand \bibinfo  [0]{\@secondoftwo}%
\providecommand \bibfield  [0]{\@secondoftwo}%
\providecommand \translation [1]{[#1]}%
\providecommand \BibitemOpen [0]{}%
\providecommand \bibitemStop [0]{}%
\providecommand \bibitemNoStop [0]{.\EOS\space}%
\providecommand \EOS [0]{\spacefactor3000\relax}%
\providecommand \BibitemShut  [1]{\csname bibitem#1\endcsname}%
\let\auto@bib@innerbib\@empty
\bibitem [{\citenamefont {Bonaccorso}\ \emph {et~al.}(2010)\citenamefont
  {Bonaccorso}, \citenamefont {Sun}, \citenamefont {Hasan},\ and\ \citenamefont
  {Ferrari}}]{bonaccorso2010}%
  \BibitemOpen
  \bibfield  {author} {\bibinfo {author} {\bibfnamefont {F.}~\bibnamefont
  {Bonaccorso}}, \bibinfo {author} {\bibfnamefont {Z.}~\bibnamefont {Sun}},
  \bibinfo {author} {\bibfnamefont {T.}~\bibnamefont {Hasan}}, \ and\ \bibinfo
  {author} {\bibfnamefont {A.~C.}\ \bibnamefont {Ferrari}},\ }\bibfield
  {title} {\enquote {\bibinfo {title} {Graphene photonics and
  optoelectronics},}\ }\href {\doibase 10.1038/nphoton.2010.186} {\bibfield
  {journal} {\bibinfo  {journal} {Nature Photonics}\ }\textbf {\bibinfo
  {volume} {4}},\ \bibinfo {pages} {611--622} (\bibinfo {year}
  {2010})}\BibitemShut {NoStop}%
\bibitem [{\citenamefont {Koppens}\ \emph {et~al.}(2014)\citenamefont
  {Koppens}, \citenamefont {Mueller}, \citenamefont {Avouris}, \citenamefont
  {Ferrari}, \citenamefont {Vitiello},\ and\ \citenamefont
  {Polini}}]{koppens2014}%
  \BibitemOpen
  \bibfield  {author} {\bibinfo {author} {\bibfnamefont {F.~H.~L.}\
  \bibnamefont {Koppens}}, \bibinfo {author} {\bibfnamefont {T.}~\bibnamefont
  {Mueller}}, \bibinfo {author} {\bibfnamefont {Ph}~\bibnamefont {Avouris}},
  \bibinfo {author} {\bibfnamefont {A.~C.}\ \bibnamefont {Ferrari}}, \bibinfo
  {author} {\bibfnamefont {M.~S.}\ \bibnamefont {Vitiello}}, \ and\ \bibinfo
  {author} {\bibfnamefont {M.}~\bibnamefont {Polini}},\ }\bibfield  {title}
  {\enquote {\bibinfo {title} {Photodetectors based on graphene, other
  two-dimensional materials and hybrid systems},}\ }\href {\doibase
  10.1038/nnano.2014.215} {\bibfield  {journal} {\bibinfo  {journal} {Nature
  Nanotechnology}\ }\textbf {\bibinfo {volume} {9}},\ \bibinfo {pages}
  {780--793} (\bibinfo {year} {2014})}\BibitemShut {NoStop}%
\bibitem [{\citenamefont {Talapin}\ \emph {et~al.}(2010)\citenamefont
  {Talapin}, \citenamefont {Lee}, \citenamefont {Kovalenko},\ and\
  \citenamefont {Shevchenko}}]{talapin2010}%
  \BibitemOpen
  \bibfield  {author} {\bibinfo {author} {\bibfnamefont {Dmitri~V.}\
  \bibnamefont {Talapin}}, \bibinfo {author} {\bibfnamefont {Jong-Soo}\
  \bibnamefont {Lee}}, \bibinfo {author} {\bibfnamefont {Maksym~V.}\
  \bibnamefont {Kovalenko}}, \ and\ \bibinfo {author} {\bibfnamefont
  {Elena~V.}\ \bibnamefont {Shevchenko}},\ }\bibfield  {title} {\enquote
  {\bibinfo {title} {Prospects of colloidal nanocrystals for electronic and
  optoelectronic applications},}\ }\href {\doibase 10.1021/cr900137k}
  {\bibfield  {journal} {\bibinfo  {journal} {Chemical Reviews}\ }\textbf
  {\bibinfo {volume} {110}},\ \bibinfo {pages} {389--458} (\bibinfo {year}
  {2010})}\BibitemShut {NoStop}%
\bibitem [{\citenamefont {Nair}\ \emph {et~al.}(2008)\citenamefont {Nair},
  \citenamefont {Blake}, \citenamefont {Grigorenko}, \citenamefont {Novoselov},
  \citenamefont {Booth}, \citenamefont {Stauber}, \citenamefont {Peres},\ and\
  \citenamefont {Geim}}]{nair2008}%
  \BibitemOpen
  \bibfield  {author} {\bibinfo {author} {\bibfnamefont {R.~R.}\ \bibnamefont
  {Nair}}, \bibinfo {author} {\bibfnamefont {P.}~\bibnamefont {Blake}},
  \bibinfo {author} {\bibfnamefont {A.~N.}\ \bibnamefont {Grigorenko}},
  \bibinfo {author} {\bibfnamefont {K.~S.}\ \bibnamefont {Novoselov}}, \bibinfo
  {author} {\bibfnamefont {T.~J.}\ \bibnamefont {Booth}}, \bibinfo {author}
  {\bibfnamefont {T.}~\bibnamefont {Stauber}}, \bibinfo {author} {\bibfnamefont
  {N.~M.~R.}\ \bibnamefont {Peres}}, \ and\ \bibinfo {author} {\bibfnamefont
  {A.~K.}\ \bibnamefont {Geim}},\ }\bibfield  {title} {\enquote {\bibinfo
  {title} {{Fine Structure Constant Defines Visual Transparency of
  Graphene}},}\ }\href {\doibase 10.1126/science.1156965} {\bibfield  {journal}
  {\bibinfo  {journal} {Science}\ }\textbf {\bibinfo {volume} {320}},\ \bibinfo
  {pages} {1308} (\bibinfo {year} {2008})}\BibitemShut {NoStop}%
\bibitem [{\citenamefont {Mak}\ \emph {et~al.}(2008)\citenamefont {Mak},
  \citenamefont {Sfeir}, \citenamefont {Wu}, \citenamefont {Lui}, \citenamefont
  {Misewich},\ and\ \citenamefont {Heinz}}]{mak2008}%
  \BibitemOpen
  \bibfield  {author} {\bibinfo {author} {\bibfnamefont {Kin~Fai}\ \bibnamefont
  {Mak}}, \bibinfo {author} {\bibfnamefont {Matthew~Y.}\ \bibnamefont {Sfeir}},
  \bibinfo {author} {\bibfnamefont {Yang}\ \bibnamefont {Wu}}, \bibinfo
  {author} {\bibfnamefont {Chun~Hung}\ \bibnamefont {Lui}}, \bibinfo {author}
  {\bibfnamefont {James~A.}\ \bibnamefont {Misewich}}, \ and\ \bibinfo {author}
  {\bibfnamefont {Tony~F.}\ \bibnamefont {Heinz}},\ }\bibfield  {title}
  {\enquote {\bibinfo {title} {Measurement of the optical conductivity of
  graphene},}\ }\href@noop {} {\bibfield  {journal} {\bibinfo  {journal}
  {Physical Review Letters}\ }\textbf {\bibinfo {volume} {101}},\ \bibinfo
  {pages} {196405} (\bibinfo {year} {2008})}\BibitemShut {NoStop}%
\bibitem [{\citenamefont {Das~Sarma}\ \emph {et~al.}(2011)\citenamefont
  {Das~Sarma}, \citenamefont {Adam}, \citenamefont {Hwang},\ and\ \citenamefont
  {Rossi}}]{dassarma2011}%
  \BibitemOpen
  \bibfield  {author} {\bibinfo {author} {\bibfnamefont {S.}~\bibnamefont
  {Das~Sarma}}, \bibinfo {author} {\bibfnamefont {Shaffique}\ \bibnamefont
  {Adam}}, \bibinfo {author} {\bibfnamefont {E.~H.}\ \bibnamefont {Hwang}}, \
  and\ \bibinfo {author} {\bibfnamefont {Enrico}\ \bibnamefont {Rossi}},\
  }\bibfield  {title} {\enquote {\bibinfo {title} {Electronic transport in
  two-dimensional graphene},}\ }\href {\doibase 10.1103/RevModPhys.83.407}
  {\bibfield  {journal} {\bibinfo  {journal} {Review of Modern Physics}\
  }\textbf {\bibinfo {volume} {83}},\ \bibinfo {pages} {407--470} (\bibinfo
  {year} {2011})}\BibitemShut {NoStop}%
\bibitem [{\citenamefont {Wang}\ \emph
  {et~al.}(2008{\natexlab{a}})\citenamefont {Wang}, \citenamefont {Zhi},\ and\
  \citenamefont {MÃ¼llen}}]{wang2008}%
  \BibitemOpen
  \bibfield  {author} {\bibinfo {author} {\bibfnamefont {Xuan}\ \bibnamefont
  {Wang}}, \bibinfo {author} {\bibfnamefont {Linjie}\ \bibnamefont {Zhi}}, \
  and\ \bibinfo {author} {\bibfnamefont {Klaus}\ \bibnamefont {MÃ¼llen}},\
  }\bibfield  {title} {\enquote {\bibinfo {title} {Transparent, conductive
  graphene electrodes for dye-sensitized solar cells},}\ }\href {\doibase
  10.1021/nl072838r} {\bibfield  {journal} {\bibinfo  {journal} {Nano Letters}\
  }\textbf {\bibinfo {volume} {8}},\ \bibinfo {pages} {323} (\bibinfo {year}
  {2008}{\natexlab{a}})}\BibitemShut {NoStop}%
\bibitem [{\citenamefont {Kim}\ \emph {et~al.}(2009{\natexlab{a}})\citenamefont
  {Kim}, \citenamefont {Zhao}, \citenamefont {Jang}, \citenamefont {Lee},
  \citenamefont {Kim}, \citenamefont {Kim}, \citenamefont {Ahn}, \citenamefont
  {Kim}, \citenamefont {Choi},\ and\ \citenamefont {Hong}}]{kim2009b}%
  \BibitemOpen
  \bibfield  {author} {\bibinfo {author} {\bibfnamefont {Keun~Soo}\
  \bibnamefont {Kim}}, \bibinfo {author} {\bibfnamefont {Yue}\ \bibnamefont
  {Zhao}}, \bibinfo {author} {\bibfnamefont {Houk}\ \bibnamefont {Jang}},
  \bibinfo {author} {\bibfnamefont {Sang~Yoon}\ \bibnamefont {Lee}}, \bibinfo
  {author} {\bibfnamefont {Jong~Min}\ \bibnamefont {Kim}}, \bibinfo {author}
  {\bibfnamefont {Kwang~S.}\ \bibnamefont {Kim}}, \bibinfo {author}
  {\bibfnamefont {Jong-Hyun}\ \bibnamefont {Ahn}}, \bibinfo {author}
  {\bibfnamefont {Philip}\ \bibnamefont {Kim}}, \bibinfo {author}
  {\bibfnamefont {Jae-Young}\ \bibnamefont {Choi}}, \ and\ \bibinfo {author}
  {\bibfnamefont {Byung~Hee}\ \bibnamefont {Hong}},\ }\bibfield  {title}
  {\enquote {\bibinfo {title} {Large-scale pattern growth of graphene films for
  stretchable transparent electrodes},}\ }\href {\doibase 10.1038/nature07719}
  {\bibfield  {journal} {\bibinfo  {journal} {Nature}\ }\textbf {\bibinfo
  {volume} {457}},\ \bibinfo {pages} {706--710} (\bibinfo {year}
  {2009}{\natexlab{a}})}\BibitemShut {NoStop}%
\bibitem [{\citenamefont {Edited~by Klimov}(2010)}]{KlimovBOOK}%
  \BibitemOpen
  \bibfield  {author} {\bibinfo {author} {\bibfnamefont {V.I.}\ \bibnamefont
  {Edited~by Klimov}},\ }\href@noop {} {\emph {\bibinfo {title} {Nanocrystal
  Quantum Dots}}}\ (\bibinfo  {publisher} {CRC Press},\ \bibinfo {address}
  {UK},\ \bibinfo {year} {2010})\BibitemShut {NoStop}%
\bibitem [{\citenamefont {Peng}\ \emph {et~al.}(2000)\citenamefont {Peng},
  \citenamefont {Manna}, \citenamefont {Yang}, \citenamefont {Wickham},
  \citenamefont {Scher}, \citenamefont {Kadavanich},\ and\ \citenamefont
  {Alivisatos}}]{peng2000}%
  \BibitemOpen
  \bibfield  {author} {\bibinfo {author} {\bibfnamefont {Xiaogang}\
  \bibnamefont {Peng}}, \bibinfo {author} {\bibfnamefont {Liberato}\
  \bibnamefont {Manna}}, \bibinfo {author} {\bibfnamefont {Weidong}\
  \bibnamefont {Yang}}, \bibinfo {author} {\bibfnamefont {Juanita}\
  \bibnamefont {Wickham}}, \bibinfo {author} {\bibfnamefont {Erik}\
  \bibnamefont {Scher}}, \bibinfo {author} {\bibfnamefont {Andreas}\
  \bibnamefont {Kadavanich}}, \ and\ \bibinfo {author} {\bibfnamefont {A.~P.}\
  \bibnamefont {Alivisatos}},\ }\bibfield  {title} {\enquote {\bibinfo {title}
  {Shape control of {CdSe} nanocrystals},}\ }\href {\doibase 10.1038/35003535}
  {\bibfield  {journal} {\bibinfo  {journal} {Nature}\ }\textbf {\bibinfo
  {volume} {404}},\ \bibinfo {pages} {59--61} (\bibinfo {year}
  {2000})}\BibitemShut {NoStop}%
\bibitem [{\citenamefont {Ithurria}\ and\ \citenamefont
  {Dubertret}(2008)}]{ithurria2008}%
  \BibitemOpen
  \bibfield  {author} {\bibinfo {author} {\bibfnamefont {Sandrine}\
  \bibnamefont {Ithurria}}\ and\ \bibinfo {author} {\bibfnamefont {Benoit}\
  \bibnamefont {Dubertret}},\ }\bibfield  {title} {\enquote {\bibinfo {title}
  {Quasi {2D} colloidal {CdSe} platelets with thicknesses controlled at the
  atomic level},}\ }\href {\doibase 10.1021/ja807724e} {\bibfield  {journal}
  {\bibinfo  {journal} {Journal of the American Chemical Society}\ }\textbf
  {\bibinfo {volume} {130}},\ \bibinfo {pages} {16504--16505} (\bibinfo {year}
  {2008})}\BibitemShut {NoStop}%
\bibitem [{\citenamefont {Ithurria}\ \emph {et~al.}(2011)\citenamefont
  {Ithurria}, \citenamefont {Tessier}, \citenamefont {Mahler}, \citenamefont
  {Lobo}, \citenamefont {Dubertret},\ and\ \citenamefont
  {Efros}}]{ithurria2011}%
  \BibitemOpen
  \bibfield  {author} {\bibinfo {author} {\bibfnamefont {S.}~\bibnamefont
  {Ithurria}}, \bibinfo {author} {\bibfnamefont {M.~D.}\ \bibnamefont
  {Tessier}}, \bibinfo {author} {\bibfnamefont {B.}~\bibnamefont {Mahler}},
  \bibinfo {author} {\bibfnamefont {R.~P. S.~M.}\ \bibnamefont {Lobo}},
  \bibinfo {author} {\bibfnamefont {B.}~\bibnamefont {Dubertret}}, \ and\
  \bibinfo {author} {\bibfnamefont {Al~L.}\ \bibnamefont {Efros}},\ }\bibfield
  {title} {\enquote {\bibinfo {title} {Colloidal nanoplatelets with
  two-dimensional electronic structure},}\ }\href {\doibase 10.1038/nmat3145}
  {\bibfield  {journal} {\bibinfo  {journal} {Nature Materials}\ }\textbf
  {\bibinfo {volume} {10}},\ \bibinfo {pages} {936--941} (\bibinfo {year}
  {2011})}\BibitemShut {NoStop}%
\bibitem [{\citenamefont {Chen}\ \emph {et~al.}(2010)\citenamefont {Chen},
  \citenamefont {Berciaud}, \citenamefont {Nuckolls}, \citenamefont {Heinz},\
  and\ \citenamefont {Brus}}]{chen2010}%
  \BibitemOpen
  \bibfield  {author} {\bibinfo {author} {\bibfnamefont {Zheyuan}\ \bibnamefont
  {Chen}}, \bibinfo {author} {\bibfnamefont {SteÌ?phane}\ \bibnamefont
  {Berciaud}}, \bibinfo {author} {\bibfnamefont {Colin}\ \bibnamefont
  {Nuckolls}}, \bibinfo {author} {\bibfnamefont {Tony~F.}\ \bibnamefont
  {Heinz}}, \ and\ \bibinfo {author} {\bibfnamefont {Louis~E.}\ \bibnamefont
  {Brus}},\ }\bibfield  {title} {\enquote {\bibinfo {title} {Energy transfer
  from individual semiconductor nanocrystals to graphene},}\ }\href {\doibase
  10.1021/nn1005107} {\bibfield  {journal} {\bibinfo  {journal} {{ACS} Nano}\
  }\textbf {\bibinfo {volume} {4}},\ \bibinfo {pages} {2964--2968} (\bibinfo
  {year} {2010})}\BibitemShut {NoStop}%
\bibitem [{\citenamefont {Ajayi}\ \emph {et~al.}(2014)\citenamefont {Ajayi},
  \citenamefont {Anderson}, \citenamefont {Cotlet}, \citenamefont {Petrone},
  \citenamefont {Gu}, \citenamefont {Wolcott}, \citenamefont {Gesuele},
  \citenamefont {Hone}, \citenamefont {Owen},\ and\ \citenamefont
  {Wong}}]{ajayi2014}%
  \BibitemOpen
  \bibfield  {author} {\bibinfo {author} {\bibfnamefont {O.~A.}\ \bibnamefont
  {Ajayi}}, \bibinfo {author} {\bibfnamefont {N.~C.}\ \bibnamefont {Anderson}},
  \bibinfo {author} {\bibfnamefont {M.}~\bibnamefont {Cotlet}}, \bibinfo
  {author} {\bibfnamefont {N.}~\bibnamefont {Petrone}}, \bibinfo {author}
  {\bibfnamefont {T.}~\bibnamefont {Gu}}, \bibinfo {author} {\bibfnamefont
  {A.}~\bibnamefont {Wolcott}}, \bibinfo {author} {\bibfnamefont
  {F.}~\bibnamefont {Gesuele}}, \bibinfo {author} {\bibfnamefont
  {J.}~\bibnamefont {Hone}}, \bibinfo {author} {\bibfnamefont {J.~S.}\
  \bibnamefont {Owen}}, \ and\ \bibinfo {author} {\bibfnamefont {C.~W.}\
  \bibnamefont {Wong}},\ }\bibfield  {title} {\enquote {\bibinfo {title}
  {Time-resolved energy transfer from single chloride-terminated nanocrystals
  to graphene},}\ }\href {\doibase 10.1063/1.4874298} {\bibfield  {journal}
  {\bibinfo  {journal} {Applied Physics Letters}\ }\textbf {\bibinfo {volume}
  {104}},\ \bibinfo {eid} {171101} (\bibinfo {year} {2014})}\BibitemShut
  {NoStop}%
\bibitem [{\citenamefont {Rogez}\ \emph {et~al.}(2014)\citenamefont {Rogez},
  \citenamefont {Yang}, \citenamefont {Le~Moal}, \citenamefont
  {L\'ev\^eque-Fort}, \citenamefont {Boer-Duchemin}, \citenamefont {Yao},
  \citenamefont {Lee}, \citenamefont {Zhang}, \citenamefont {Wegner},
  \citenamefont {Hildebrandt}, \citenamefont {Mayne},\ and\ \citenamefont
  {Dujardin}}]{rogez2014}%
  \BibitemOpen
  \bibfield  {author} {\bibinfo {author} {\bibfnamefont {Beno\"it}\
  \bibnamefont {Rogez}}, \bibinfo {author} {\bibfnamefont {Heejun}\
  \bibnamefont {Yang}}, \bibinfo {author} {\bibfnamefont {Eric}\ \bibnamefont
  {Le~Moal}}, \bibinfo {author} {\bibfnamefont {Sandrine}\ \bibnamefont
  {L\'ev\^eque-Fort}}, \bibinfo {author} {\bibfnamefont {Elizabeth}\
  \bibnamefont {Boer-Duchemin}}, \bibinfo {author} {\bibfnamefont {Fei}\
  \bibnamefont {Yao}}, \bibinfo {author} {\bibfnamefont {Young-Hee}\
  \bibnamefont {Lee}}, \bibinfo {author} {\bibfnamefont {Yang}\ \bibnamefont
  {Zhang}}, \bibinfo {author} {\bibfnamefont {K.~David}\ \bibnamefont
  {Wegner}}, \bibinfo {author} {\bibfnamefont {Niko}\ \bibnamefont
  {Hildebrandt}}, \bibinfo {author} {\bibfnamefont {Andrew}\ \bibnamefont
  {Mayne}}, \ and\ \bibinfo {author} {\bibfnamefont {G\'erald}\ \bibnamefont
  {Dujardin}},\ }\bibfield  {title} {\enquote {\bibinfo {title} {Fluorescence
  lifetime and blinking of individual semiconductor nanocrystals on
  graphene},}\ }\href {\doibase 10.1021/jp5061446} {\bibfield  {journal}
  {\bibinfo  {journal} {The Journal of Physical Chemistry C}\ }\textbf
  {\bibinfo {volume} {118}},\ \bibinfo {pages} {18445--18452} (\bibinfo {year}
  {2014})}\BibitemShut {NoStop}%
\bibitem [{\citenamefont {Konstantatos}\ \emph {et~al.}(2012)\citenamefont
  {Konstantatos}, \citenamefont {Badioli}, \citenamefont {Gaudreau},
  \citenamefont {Osmond}, \citenamefont {Bernechea}, \citenamefont {de~Arquer},
  \citenamefont {Gatti},\ and\ \citenamefont {Koppens}}]{konstantatos2012}%
  \BibitemOpen
  \bibfield  {author} {\bibinfo {author} {\bibfnamefont {Gerasimos}\
  \bibnamefont {Konstantatos}}, \bibinfo {author} {\bibfnamefont {Michela}\
  \bibnamefont {Badioli}}, \bibinfo {author} {\bibfnamefont {Louis}\
  \bibnamefont {Gaudreau}}, \bibinfo {author} {\bibfnamefont {Johann}\
  \bibnamefont {Osmond}}, \bibinfo {author} {\bibfnamefont {Maria}\
  \bibnamefont {Bernechea}}, \bibinfo {author} {\bibfnamefont
  {F.~Pelayo~Garcia}\ \bibnamefont {de~Arquer}}, \bibinfo {author}
  {\bibfnamefont {Fabio}\ \bibnamefont {Gatti}}, \ and\ \bibinfo {author}
  {\bibfnamefont {Frank H.~L.}\ \bibnamefont {Koppens}},\ }\bibfield  {title}
  {\enquote {\bibinfo {title} {Hybrid graphene-quantum dot phototransistors
  with ultrahigh gain},}\ }\href {\doibase 10.1038/nnano.2012.60} {\bibfield
  {journal} {\bibinfo  {journal} {Nature Nanotechnology}\ }\textbf {\bibinfo
  {volume} {7}},\ \bibinfo {pages} {363} (\bibinfo {year} {2012})}\BibitemShut
  {NoStop}%
\bibitem [{\citenamefont {Sun}\ \emph {et~al.}(2012)\citenamefont {Sun},
  \citenamefont {Liu}, \citenamefont {Li}, \citenamefont {Tai}, \citenamefont
  {Lau},\ and\ \citenamefont {Yan}}]{sun2012}%
  \BibitemOpen
  \bibfield  {author} {\bibinfo {author} {\bibfnamefont {Zhenhua}\ \bibnamefont
  {Sun}}, \bibinfo {author} {\bibfnamefont {Zhike}\ \bibnamefont {Liu}},
  \bibinfo {author} {\bibfnamefont {Jinhua}\ \bibnamefont {Li}}, \bibinfo
  {author} {\bibfnamefont {Guo-an}\ \bibnamefont {Tai}}, \bibinfo {author}
  {\bibfnamefont {Shu-Ping}\ \bibnamefont {Lau}}, \ and\ \bibinfo {author}
  {\bibfnamefont {Feng}\ \bibnamefont {Yan}},\ }\bibfield  {title} {\enquote
  {\bibinfo {title} {Infrared photodetectors based on cvd-grown graphene and
  pbs quantum dots with ultrahigh responsivity},}\ }\href {\doibase
  10.1002/adma.201202220} {\bibfield  {journal} {\bibinfo  {journal} {Advanced
  Materials}\ }\textbf {\bibinfo {volume} {24}},\ \bibinfo {pages} {5878}
  (\bibinfo {year} {2012})}\BibitemShut {NoStop}%
\bibitem [{\citenamefont {Klekachev}\ \emph {et~al.}(2012)\citenamefont
  {Klekachev}, \citenamefont {Asselberghs}, \citenamefont {Kuznetsov},
  \citenamefont {Cantoro}, \citenamefont {Mun}, \citenamefont {Cho},
  \citenamefont {Hotta}, \citenamefont {Hofkens}, \citenamefont {van~der Veen},
  \citenamefont {Stesmans}, \citenamefont {Heyns},\ and\ \citenamefont
  {De~Gendt}}]{klekachev2012}%
  \BibitemOpen
  \bibfield  {author} {\bibinfo {author} {\bibfnamefont {Alexander~V.}\
  \bibnamefont {Klekachev}}, \bibinfo {author} {\bibfnamefont {Inge}\
  \bibnamefont {Asselberghs}}, \bibinfo {author} {\bibfnamefont {Sergey~N.}\
  \bibnamefont {Kuznetsov}}, \bibinfo {author} {\bibfnamefont {Mirco}\
  \bibnamefont {Cantoro}}, \bibinfo {author} {\bibfnamefont {Jeong~Hun}\
  \bibnamefont {Mun}}, \bibinfo {author} {\bibfnamefont {Byung-Jin}\
  \bibnamefont {Cho}}, \bibinfo {author} {\bibfnamefont {Jun-ichi}\
  \bibnamefont {Hotta}}, \bibinfo {author} {\bibfnamefont {Johan}\ \bibnamefont
  {Hofkens}}, \bibinfo {author} {\bibfnamefont {Marleen}\ \bibnamefont {van~der
  Veen}}, \bibinfo {author} {\bibfnamefont {André~L.}\ \bibnamefont
  {Stesmans}}, \bibinfo {author} {\bibfnamefont {Marc~M.}\ \bibnamefont
  {Heyns}}, \ and\ \bibinfo {author} {\bibfnamefont {Stefan}\ \bibnamefont
  {De~Gendt}},\ }\bibfield  {title} {\enquote {\bibinfo {title} {Charge
  transfer effects in graphene-cdse/zns quantum dots composites},}\ }\href
  {\doibase 10.1117/12.930082} {\bibfield  {journal} {\bibinfo  {journal}
  {Proc. SPIE}\ }\textbf {\bibinfo {volume} {8462}},\ \bibinfo {pages} {84620L}
  (\bibinfo {year} {2012})}\BibitemShut {NoStop}%
\bibitem [{\citenamefont {Klekachev}\ \emph {et~al.}(2013)\citenamefont
  {Klekachev}, \citenamefont {Kuznetsov}, \citenamefont {Asselberghs},
  \citenamefont {Cantoro}, \citenamefont {Hun~Mun}, \citenamefont {Jin~Cho},
  \citenamefont {Stesmans}, \citenamefont {Heyns},\ and\ \citenamefont
  {De~Gendt}}]{klekachev2013}%
  \BibitemOpen
  \bibfield  {author} {\bibinfo {author} {\bibfnamefont {Alexander~V.}\
  \bibnamefont {Klekachev}}, \bibinfo {author} {\bibfnamefont {Sergey~N.}\
  \bibnamefont {Kuznetsov}}, \bibinfo {author} {\bibfnamefont {Inge}\
  \bibnamefont {Asselberghs}}, \bibinfo {author} {\bibfnamefont {Mirco}\
  \bibnamefont {Cantoro}}, \bibinfo {author} {\bibfnamefont {Jeong}\
  \bibnamefont {Hun~Mun}}, \bibinfo {author} {\bibfnamefont {Byung}\
  \bibnamefont {Jin~Cho}}, \bibinfo {author} {\bibfnamefont {Andr\'e~L.}\
  \bibnamefont {Stesmans}}, \bibinfo {author} {\bibfnamefont {Marc~M.}\
  \bibnamefont {Heyns}}, \ and\ \bibinfo {author} {\bibfnamefont {Stefan}\
  \bibnamefont {De~Gendt}},\ }\bibfield  {title} {\enquote {\bibinfo {title}
  {Graphene as anode electrode for colloidal quantum dots based light emitting
  diodes},}\ }\href {\doibase 10.1063/1.4816745} {\bibfield  {journal}
  {\bibinfo  {journal} {Applied Physics Letters}\ }\textbf {\bibinfo {volume}
  {103}},\ \bibinfo {eid} {043124} (\bibinfo {year} {2013})}\BibitemShut
  {NoStop}%
\bibitem [{\citenamefont {F\"orster}(1948)}]{forster1948}%
  \BibitemOpen
  \bibfield  {author} {\bibinfo {author} {\bibfnamefont {Th}~\bibnamefont
  {F\"orster}},\ }\bibfield  {title} {\enquote {\bibinfo {title}
  {Zwischenmolekulare energiewanderung und fluoreszenz},}\ }\href
  {http://onlinelibrary.wiley.com/doi/10.1002/andp.19484370105/abstract}
  {\bibfield  {journal} {\bibinfo  {journal} {Annalen der physik}\ }\textbf
  {\bibinfo {volume} {437}},\ \bibinfo {pages} {55} (\bibinfo {year}
  {1948})}\BibitemShut {NoStop}%
\bibitem [{\citenamefont {Velizhanin}\ and\ \citenamefont
  {Efimov}(2011)}]{velizhanin2011}%
  \BibitemOpen
  \bibfield  {author} {\bibinfo {author} {\bibfnamefont {Kirill~A.}\
  \bibnamefont {Velizhanin}}\ and\ \bibinfo {author} {\bibfnamefont {Anatoly}\
  \bibnamefont {Efimov}},\ }\bibfield  {title} {\enquote {\bibinfo {title}
  {Probing plasmons in graphene by resonance energy transfer},}\ }\href
  {\doibase 10.1103/PhysRevB.84.085401} {\bibfield  {journal} {\bibinfo
  {journal} {Phys. Rev. B}\ }\textbf {\bibinfo {volume} {84}},\ \bibinfo
  {pages} {085401} (\bibinfo {year} {2011})}\BibitemShut {NoStop}%
\bibitem [{\citenamefont {GÃ³mez-Santos}\ and\ \citenamefont
  {Stauber}(2011)}]{gomez2011}%
  \BibitemOpen
  \bibfield  {author} {\bibinfo {author} {\bibfnamefont {G.}~\bibnamefont
  {GÃ³mez-Santos}}\ and\ \bibinfo {author} {\bibfnamefont {T.}~\bibnamefont
  {Stauber}},\ }\bibfield  {title} {\enquote {\bibinfo {title} {Fluorescence
  quenching in graphene: A fundamental ruler and evidence for transverse
  plasmons},}\ }\href {\doibase 10.1103/PhysRevB.84.165438} {\bibfield
  {journal} {\bibinfo  {journal} {Phys. Rev. B}\ }\textbf {\bibinfo {volume}
  {84}},\ \bibinfo {pages} {165438} (\bibinfo {year} {2011})}\BibitemShut
  {NoStop}%
\bibitem [{\citenamefont {Tielrooij}\ \emph {et~al.}(2014)\citenamefont
  {Tielrooij}, \citenamefont {Orona}, \citenamefont {Ferrier}, \citenamefont
  {Badioli}, \citenamefont {Navickaite}, \citenamefont {Coop}, \citenamefont
  {Nanot}, \citenamefont {Kalinic}, \citenamefont {Cesca}, \citenamefont
  {Gaudreau}, \citenamefont {Ma}, \citenamefont {Centeno}, \citenamefont
  {Pesquera}, \citenamefont {Zurutuza}, \citenamefont {de~Riedmatten},
  \citenamefont {Goldner}, \citenamefont {de~Abajo}, \citenamefont
  {Jarillo-Herrero},\ and\ \citenamefont {Koppens}}]{tielrooij2014}%
  \BibitemOpen
  \bibfield  {author} {\bibinfo {author} {\bibfnamefont {K.~J.}\ \bibnamefont
  {Tielrooij}}, \bibinfo {author} {\bibfnamefont {L.}~\bibnamefont {Orona}},
  \bibinfo {author} {\bibfnamefont {A.}~\bibnamefont {Ferrier}}, \bibinfo
  {author} {\bibfnamefont {M.}~\bibnamefont {Badioli}}, \bibinfo {author}
  {\bibfnamefont {G.}~\bibnamefont {Navickaite}}, \bibinfo {author}
  {\bibfnamefont {S.}~\bibnamefont {Coop}}, \bibinfo {author} {\bibfnamefont
  {S.}~\bibnamefont {Nanot}}, \bibinfo {author} {\bibfnamefont
  {B.}~\bibnamefont {Kalinic}}, \bibinfo {author} {\bibfnamefont
  {T.}~\bibnamefont {Cesca}}, \bibinfo {author} {\bibfnamefont
  {L.}~\bibnamefont {Gaudreau}}, \bibinfo {author} {\bibfnamefont
  {Q.}~\bibnamefont {Ma}}, \bibinfo {author} {\bibfnamefont {A.}~\bibnamefont
  {Centeno}}, \bibinfo {author} {\bibfnamefont {A.}~\bibnamefont {Pesquera}},
  \bibinfo {author} {\bibfnamefont {A.}~\bibnamefont {Zurutuza}}, \bibinfo
  {author} {\bibfnamefont {H.}~\bibnamefont {de~Riedmatten}}, \bibinfo {author}
  {\bibfnamefont {P.}~\bibnamefont {Goldner}}, \bibinfo {author} {\bibfnamefont
  {F.~J.~Garcia}\ \bibnamefont {de~Abajo}}, \bibinfo {author} {\bibfnamefont
  {P.}~\bibnamefont {Jarillo-Herrero}}, \ and\ \bibinfo {author} {\bibfnamefont
  {F.~H.~L.}\ \bibnamefont {Koppens}},\ }\bibfield  {title} {\enquote {\bibinfo
  {title} {Electrical control of optical emitter relaxation pathways enabled by
  graphene},}\ }\href {http://arxiv.org/abs/1410.1361} {\bibfield  {journal}
  {\bibinfo  {journal} {{arXiv}:1410.1361}\ } (\bibinfo {year}
  {2014})}\BibitemShut {NoStop}%
\bibitem [{\citenamefont {Lee}\ \emph {et~al.}(2014)\citenamefont {Lee},
  \citenamefont {Bao}, \citenamefont {Ju}, \citenamefont {Schuck},
  \citenamefont {Wang},\ and\ \citenamefont {Weber-Bargioni}}]{lee2014}%
  \BibitemOpen
  \bibfield  {author} {\bibinfo {author} {\bibfnamefont {Jiye}\ \bibnamefont
  {Lee}}, \bibinfo {author} {\bibfnamefont {Wei}\ \bibnamefont {Bao}}, \bibinfo
  {author} {\bibfnamefont {Long}\ \bibnamefont {Ju}}, \bibinfo {author}
  {\bibfnamefont {P.~James}\ \bibnamefont {Schuck}}, \bibinfo {author}
  {\bibfnamefont {Feng}\ \bibnamefont {Wang}}, \ and\ \bibinfo {author}
  {\bibfnamefont {Alexander}\ \bibnamefont {Weber-Bargioni}},\ }\bibfield
  {title} {\enquote {\bibinfo {title} {Switching individual quantum dot
  emission through electrically controlling resonant energy transfer to
  graphene},}\ }\href {\doibase 10.1021/nl503587z} {\bibfield  {journal}
  {\bibinfo  {journal} {Nano Letters}\ }\textbf {\bibinfo {volume} {14}},\
  \bibinfo {pages} {7115--7119} (\bibinfo {year} {2014})}\BibitemShut {NoStop}%
\bibitem [{\citenamefont {Shafran}\ \emph {et~al.}(2010)\citenamefont
  {Shafran}, \citenamefont {Mangum},\ and\ \citenamefont
  {Gerton}}]{shafran2010}%
  \BibitemOpen
  \bibfield  {author} {\bibinfo {author} {\bibfnamefont {Eyal}\ \bibnamefont
  {Shafran}}, \bibinfo {author} {\bibfnamefont {Benjamin~D.}\ \bibnamefont
  {Mangum}}, \ and\ \bibinfo {author} {\bibfnamefont {Jordan~M.}\ \bibnamefont
  {Gerton}},\ }\bibfield  {title} {\enquote {\bibinfo {title} {Energy transfer
  from an individual quantum dot to a carbon nanotube},}\ }\href {\doibase
  10.1021/nl102045g} {\bibfield  {journal} {\bibinfo  {journal} {Nano Letters}\
  }\textbf {\bibinfo {volume} {10}},\ \bibinfo {pages} {4049--4054} (\bibinfo
  {year} {2010})}\BibitemShut {NoStop}%
\bibitem [{\citenamefont {Jander}\ \emph {et~al.}(2011)\citenamefont {Jander},
  \citenamefont {Kornowski},\ and\ \citenamefont {Weller}}]{jander2011}%
  \BibitemOpen
  \bibfield  {author} {\bibinfo {author} {\bibfnamefont {Sebastian}\
  \bibnamefont {Jander}}, \bibinfo {author} {\bibfnamefont {Andreas}\
  \bibnamefont {Kornowski}}, \ and\ \bibinfo {author} {\bibfnamefont {Horst}\
  \bibnamefont {Weller}},\ }\bibfield  {title} {\enquote {\bibinfo {title}
  {Energy transfer from {CdSe}/{CdS} nanorods to amorphous carbon},}\ }\href
  {\doibase 10.1021/nl202370q} {\bibfield  {journal} {\bibinfo  {journal} {Nano
  Letters}\ }\textbf {\bibinfo {volume} {11}},\ \bibinfo {pages} {5179--5183}
  (\bibinfo {year} {2011})}\BibitemShut {NoStop}%
\bibitem [{\citenamefont {Lin}\ \emph {et~al.}(2013)\citenamefont {Lin},
  \citenamefont {Huang}, \citenamefont {Shu}, \citenamefont {Yuan},
  \citenamefont {Shen}, \citenamefont {Lin}, \citenamefont {Chang},
  \citenamefont {Chiu}, \citenamefont {Lin}, \citenamefont {Lin},\ and\
  \citenamefont {Kuo}}]{lin2013}%
  \BibitemOpen
  \bibfield  {author} {\bibinfo {author} {\bibfnamefont {T.~N.}\ \bibnamefont
  {Lin}}, \bibinfo {author} {\bibfnamefont {L.~T.}\ \bibnamefont {Huang}},
  \bibinfo {author} {\bibfnamefont {G.~W.}\ \bibnamefont {Shu}}, \bibinfo
  {author} {\bibfnamefont {C.~T.}\ \bibnamefont {Yuan}}, \bibinfo {author}
  {\bibfnamefont {J.~L.}\ \bibnamefont {Shen}}, \bibinfo {author}
  {\bibfnamefont {C.~A.~J.}\ \bibnamefont {Lin}}, \bibinfo {author}
  {\bibfnamefont {W.~H.}\ \bibnamefont {Chang}}, \bibinfo {author}
  {\bibfnamefont {C.~H.}\ \bibnamefont {Chiu}}, \bibinfo {author}
  {\bibfnamefont {D.~W.}\ \bibnamefont {Lin}}, \bibinfo {author} {\bibfnamefont
  {C.~C.}\ \bibnamefont {Lin}}, \ and\ \bibinfo {author} {\bibfnamefont
  {H.~C.}\ \bibnamefont {Kuo}},\ }\bibfield  {title} {\enquote {\bibinfo
  {title} {Distance dependence of energy transfer from ingan quantum wells to
  graphene oxide},}\ }\href {\doibase 10.1364/OL.38.002897} {\bibfield
  {journal} {\bibinfo  {journal} {Opt. Lett.}\ }\textbf {\bibinfo {volume}
  {38}},\ \bibinfo {pages} {2897--2899} (\bibinfo {year} {2013})}\BibitemShut
  {NoStop}%
\bibitem [{\citenamefont {Treossi}\ \emph {et~al.}(2009)\citenamefont
  {Treossi}, \citenamefont {Melucci}, \citenamefont {Liscio}, \citenamefont
  {Gazzano}, \citenamefont {Samori},\ and\ \citenamefont
  {Palermo}}]{treossi2009}%
  \BibitemOpen
  \bibfield  {author} {\bibinfo {author} {\bibfnamefont {Emanuele}\
  \bibnamefont {Treossi}}, \bibinfo {author} {\bibfnamefont {Manuela}\
  \bibnamefont {Melucci}}, \bibinfo {author} {\bibfnamefont {Andrea}\
  \bibnamefont {Liscio}}, \bibinfo {author} {\bibfnamefont {Massimo}\
  \bibnamefont {Gazzano}}, \bibinfo {author} {\bibfnamefont {Paolo}\
  \bibnamefont {Samori}}, \ and\ \bibinfo {author} {\bibfnamefont {Vincenzo}\
  \bibnamefont {Palermo}},\ }\bibfield  {title} {\enquote {\bibinfo {title}
  {High-contrast visualization of graphene oxide on dye-sensitized glass,
  quartz, and silicon by fluorescence quenching},}\ }\href {\doibase
  10.1021/ja9055382} {\bibfield  {journal} {\bibinfo  {journal} {Journal of the
  American Chemical Society}\ }\textbf {\bibinfo {volume} {131}},\ \bibinfo
  {pages} {15576--15577} (\bibinfo {year} {2009})}\BibitemShut {NoStop}%
\bibitem [{\citenamefont {Kim}\ \emph {et~al.}(2009{\natexlab{b}})\citenamefont
  {Kim}, \citenamefont {Cote}, \citenamefont {Kim},\ and\ \citenamefont
  {Huang}}]{kim2009}%
  \BibitemOpen
  \bibfield  {author} {\bibinfo {author} {\bibfnamefont {Jaemyung}\
  \bibnamefont {Kim}}, \bibinfo {author} {\bibfnamefont {Laura~J.}\
  \bibnamefont {Cote}}, \bibinfo {author} {\bibfnamefont {Franklin}\
  \bibnamefont {Kim}}, \ and\ \bibinfo {author} {\bibfnamefont {Jiaxing}\
  \bibnamefont {Huang}},\ }\bibfield  {title} {\enquote {\bibinfo {title}
  {Visualizing graphene based sheets by fluorescence quenching microscopy},}\
  }\href {http://pubs.acs.org/doi/abs/10.1021/ja906730d} {\bibfield  {journal}
  {\bibinfo  {journal} {Journal of the American Chemical Society}\ }\textbf
  {\bibinfo {volume} {132}},\ \bibinfo {pages} {260â€“267} (\bibinfo {year}
  {2009}{\natexlab{b}})}\BibitemShut {NoStop}%
\bibitem [{\citenamefont {Gaudreau}\ \emph {et~al.}(2013)\citenamefont
  {Gaudreau}, \citenamefont {Tielrooij}, \citenamefont {Prawiroatmodjo},
  \citenamefont {Osmond}, \citenamefont {de~Abajo},\ and\ \citenamefont
  {Koppens}}]{gaudreau2013}%
  \BibitemOpen
  \bibfield  {author} {\bibinfo {author} {\bibfnamefont {L.}~\bibnamefont
  {Gaudreau}}, \bibinfo {author} {\bibfnamefont {K.~J.}\ \bibnamefont
  {Tielrooij}}, \bibinfo {author} {\bibfnamefont {G.~E. D.~K.}\ \bibnamefont
  {Prawiroatmodjo}}, \bibinfo {author} {\bibfnamefont {J.}~\bibnamefont
  {Osmond}}, \bibinfo {author} {\bibfnamefont {F.~J.~Garcia}\ \bibnamefont
  {de~Abajo}}, \ and\ \bibinfo {author} {\bibfnamefont {F.~H.~L.}\ \bibnamefont
  {Koppens}},\ }\bibfield  {title} {\enquote {\bibinfo {title} {Universal
  distance-scaling of nonradiative energy transfer to graphene},}\ }\href
  {\doibase 10.1021/nl400176b} {\bibfield  {journal} {\bibinfo  {journal} {Nano
  Letters}\ }\textbf {\bibinfo {volume} {13}},\ \bibinfo {pages} {2030--2035}
  (\bibinfo {year} {2013})}\BibitemShut {NoStop}%
\bibitem [{\citenamefont {St\"ohr}\ \emph {et~al.}(2012)\citenamefont
  {St\"ohr}, \citenamefont {Kolesov}, \citenamefont {Xia}, \citenamefont
  {Reuter}, \citenamefont {Meijer}, \citenamefont {Logvenov},\ and\
  \citenamefont {Wrachtrup}}]{stohr2012}%
  \BibitemOpen
  \bibfield  {author} {\bibinfo {author} {\bibfnamefont {Rainer~J.}\
  \bibnamefont {St\"ohr}}, \bibinfo {author} {\bibfnamefont {Roman}\
  \bibnamefont {Kolesov}}, \bibinfo {author} {\bibfnamefont {Kangwei}\
  \bibnamefont {Xia}}, \bibinfo {author} {\bibfnamefont {Rolf}\ \bibnamefont
  {Reuter}}, \bibinfo {author} {\bibfnamefont {Jan}\ \bibnamefont {Meijer}},
  \bibinfo {author} {\bibfnamefont {Gennady}\ \bibnamefont {Logvenov}}, \ and\
  \bibinfo {author} {\bibfnamefont {J\"org}\ \bibnamefont {Wrachtrup}},\
  }\bibfield  {title} {\enquote {\bibinfo {title} {Super-resolution
  fluorescence quenching microscopy of graphene},}\ }\href {\doibase
  10.1021/nn303510p} {\bibfield  {journal} {\bibinfo  {journal} {{ACS} Nano}\
  }\textbf {\bibinfo {volume} {6}},\ \bibinfo {pages} {9175--9181} (\bibinfo
  {year} {2012})}\BibitemShut {NoStop}%
\bibitem [{\citenamefont {Tisler}\ \emph {et~al.}(2013)\citenamefont {Tisler},
  \citenamefont {Oeckinghaus}, \citenamefont {St\"ohr}, \citenamefont
  {Kolesov}, \citenamefont {Reuter}, \citenamefont {Reinhard},\ and\
  \citenamefont {Wrachtrup}}]{tisler2013}%
  \BibitemOpen
  \bibfield  {author} {\bibinfo {author} {\bibfnamefont {Julia}\ \bibnamefont
  {Tisler}}, \bibinfo {author} {\bibfnamefont {Thomas}\ \bibnamefont
  {Oeckinghaus}}, \bibinfo {author} {\bibfnamefont {Rainer~J.}\ \bibnamefont
  {St\"ohr}}, \bibinfo {author} {\bibfnamefont {Roman}\ \bibnamefont
  {Kolesov}}, \bibinfo {author} {\bibfnamefont {Rolf}\ \bibnamefont {Reuter}},
  \bibinfo {author} {\bibfnamefont {Friedemann}\ \bibnamefont {Reinhard}}, \
  and\ \bibinfo {author} {\bibfnamefont {J\"org}\ \bibnamefont {Wrachtrup}},\
  }\bibfield  {title} {\enquote {\bibinfo {title} {Single defect center
  scanning near-field optical microscopy on graphene},}\ }\href {\doibase
  10.1021/nl401129m} {\bibfield  {journal} {\bibinfo  {journal} {Nano Letters}\
  }\textbf {\bibinfo {volume} {13}},\ \bibinfo {pages} {3152--3156} (\bibinfo
  {year} {2013})}\BibitemShut {NoStop}%
\bibitem [{\citenamefont {Wang}\ \emph {et~al.}(2011)\citenamefont {Wang},
  \citenamefont {Li}, \citenamefont {Wang}, \citenamefont {Li},\ and\
  \citenamefont {Lin}}]{wang2011}%
  \BibitemOpen
  \bibfield  {author} {\bibinfo {author} {\bibfnamefont {Ying}\ \bibnamefont
  {Wang}}, \bibinfo {author} {\bibfnamefont {Zhaohui}\ \bibnamefont {Li}},
  \bibinfo {author} {\bibfnamefont {Jun}\ \bibnamefont {Wang}}, \bibinfo
  {author} {\bibfnamefont {Jinghong}\ \bibnamefont {Li}}, \ and\ \bibinfo
  {author} {\bibfnamefont {Yuehe}\ \bibnamefont {Lin}},\ }\bibfield  {title}
  {\enquote {\bibinfo {title} {Graphene and graphene oxide:
  biofunctionalization and applications in biotechnology},}\ }\href {\doibase
  10.1016/j.tibtech.2011.01.008} {\bibfield  {journal} {\bibinfo  {journal}
  {Trends in Biotechnology}\ }\textbf {\bibinfo {volume} {29}},\ \bibinfo
  {pages} {205 -- 212} (\bibinfo {year} {2011})}\BibitemShut {NoStop}%
\bibitem [{\citenamefont {Mazzamuto}\ \emph {et~al.}(2014)\citenamefont
  {Mazzamuto}, \citenamefont {Tabani}, \citenamefont {Pazzagli}, \citenamefont
  {Rizvi}, \citenamefont {Reserbat-Plantey}, \citenamefont {Sch\"adler},
  \citenamefont {Navickaite}, \citenamefont {Gaudreau}, \citenamefont
  {Cataliotti}, \citenamefont {Koppens},\ and\ \citenamefont
  {Toninelli}}]{mazzamuto2014}%
  \BibitemOpen
  \bibfield  {author} {\bibinfo {author} {\bibfnamefont {G}~\bibnamefont
  {Mazzamuto}}, \bibinfo {author} {\bibfnamefont {A}~\bibnamefont {Tabani}},
  \bibinfo {author} {\bibfnamefont {S}~\bibnamefont {Pazzagli}}, \bibinfo
  {author} {\bibfnamefont {S}~\bibnamefont {Rizvi}}, \bibinfo {author}
  {\bibfnamefont {A}~\bibnamefont {Reserbat-Plantey}}, \bibinfo {author}
  {\bibfnamefont {K}~\bibnamefont {Sch\"adler}}, \bibinfo {author}
  {\bibfnamefont {G}~\bibnamefont {Navickaite}}, \bibinfo {author}
  {\bibfnamefont {L}~\bibnamefont {Gaudreau}}, \bibinfo {author} {\bibfnamefont
  {F~S}\ \bibnamefont {Cataliotti}}, \bibinfo {author} {\bibfnamefont
  {F}~\bibnamefont {Koppens}}, \ and\ \bibinfo {author} {\bibfnamefont
  {C}~\bibnamefont {Toninelli}},\ }\bibfield  {title} {\enquote {\bibinfo
  {title} {Single-molecule study for a graphene-based nano-position sensor},}\
  }\href {http://stacks.iop.org/1367-2630/16/i=11/a=113007} {\bibfield
  {journal} {\bibinfo  {journal} {New Journal of Physics}\ }\textbf {\bibinfo
  {volume} {16}},\ \bibinfo {pages} {113007} (\bibinfo {year}
  {2014})}\BibitemShut {NoStop}%
\bibitem [{\citenamefont {Halivni}\ \emph {et~al.}(2012)\citenamefont
  {Halivni}, \citenamefont {Sitt}, \citenamefont {Hadar},\ and\ \citenamefont
  {Banin}}]{halivni2012}%
  \BibitemOpen
  \bibfield  {author} {\bibinfo {author} {\bibfnamefont {Shira}\ \bibnamefont
  {Halivni}}, \bibinfo {author} {\bibfnamefont {Amit}\ \bibnamefont {Sitt}},
  \bibinfo {author} {\bibfnamefont {Ido}\ \bibnamefont {Hadar}}, \ and\
  \bibinfo {author} {\bibfnamefont {Uri}\ \bibnamefont {Banin}},\ }\bibfield
  {title} {\enquote {\bibinfo {title} {Effect of nanoparticle dimensionality on
  fluorescence resonance energy transfer in nanoparticle-dye conjugated
  systems},}\ }\href {\doibase 10.1021/nn300216v} {\bibfield  {journal}
  {\bibinfo  {journal} {{ACS} Nano}\ }\textbf {\bibinfo {volume} {6}},\
  \bibinfo {pages} {2758--2765} (\bibinfo {year} {2012})}\BibitemShut {NoStop}%
\bibitem [{\citenamefont {Rindermann}\ \emph {et~al.}(2011)\citenamefont
  {Rindermann}, \citenamefont {Pozina}, \citenamefont {Monemar}, \citenamefont
  {Hultman}, \citenamefont {Amano},\ and\ \citenamefont
  {Lagoudakis}}]{rindermann2011}%
  \BibitemOpen
  \bibfield  {author} {\bibinfo {author} {\bibfnamefont {Jan~Junis}\
  \bibnamefont {Rindermann}}, \bibinfo {author} {\bibfnamefont {Galia}\
  \bibnamefont {Pozina}}, \bibinfo {author} {\bibfnamefont {Bo}~\bibnamefont
  {Monemar}}, \bibinfo {author} {\bibfnamefont {Lars}\ \bibnamefont {Hultman}},
  \bibinfo {author} {\bibfnamefont {Hiroshi}\ \bibnamefont {Amano}}, \ and\
  \bibinfo {author} {\bibfnamefont {Pavlos~G.}\ \bibnamefont {Lagoudakis}},\
  }\bibfield  {title} {\enquote {\bibinfo {title} {Dependence of resonance
  energy transfer on exciton dimensionality},}\ }\href {\doibase
  10.1103/PhysRevLett.107.236805} {\bibfield  {journal} {\bibinfo  {journal}
  {Physical Review Letters}\ }\textbf {\bibinfo {volume} {107}},\ \bibinfo
  {pages} {236805} (\bibinfo {year} {2011})}\BibitemShut {NoStop}%
\bibitem [{\citenamefont {Godel}\ \emph {et~al.}(2013)\citenamefont {Godel},
  \citenamefont {Pichonat}, \citenamefont {Vignaud}, \citenamefont {Majjad},
  \citenamefont {Metten}, \citenamefont {Henry}, \citenamefont {Berciaud},
  \citenamefont {Dayen},\ and\ \citenamefont {Halley}}]{godel2013}%
  \BibitemOpen
  \bibfield  {author} {\bibinfo {author} {\bibfnamefont {Florian}\ \bibnamefont
  {Godel}}, \bibinfo {author} {\bibfnamefont {Emmanuelle}\ \bibnamefont
  {Pichonat}}, \bibinfo {author} {\bibfnamefont {Dominique}\ \bibnamefont
  {Vignaud}}, \bibinfo {author} {\bibfnamefont {Hicham}\ \bibnamefont
  {Majjad}}, \bibinfo {author} {\bibfnamefont {Dominik}\ \bibnamefont
  {Metten}}, \bibinfo {author} {\bibfnamefont {Yves}\ \bibnamefont {Henry}},
  \bibinfo {author} {\bibfnamefont {StÃ©phane}\ \bibnamefont {Berciaud}},
  \bibinfo {author} {\bibfnamefont {Jean-Francois}\ \bibnamefont {Dayen}}, \
  and\ \bibinfo {author} {\bibfnamefont {David}\ \bibnamefont {Halley}},\
  }\bibfield  {title} {\enquote {\bibinfo {title} {Epitaxy of {MgO} magnetic
  tunnel barriers on epitaxial graphene},}\ }\href {\doibase
  10.1088/0957-4484/24/47/475708} {\bibfield  {journal} {\bibinfo  {journal}
  {Nanotechnology}\ }\textbf {\bibinfo {volume} {24}},\ \bibinfo {pages}
  {475708} (\bibinfo {year} {2013})}\BibitemShut {NoStop}%
\bibitem [{\citenamefont {Kuhn}(1970)}]{kuhn1970}%
  \BibitemOpen
  \bibfield  {author} {\bibinfo {author} {\bibfnamefont {Hans}\ \bibnamefont
  {Kuhn}},\ }\bibfield  {title} {\enquote {\bibinfo {title} {Classical aspects
  of energy transfer in molecular systems},}\ }\href {\doibase
  10.1063/1.1673749} {\bibfield  {journal} {\bibinfo  {journal} {The Journal of
  Chemical Physics}\ }\textbf {\bibinfo {volume} {53}},\ \bibinfo {pages}
  {101--108} (\bibinfo {year} {1970})}\BibitemShut {NoStop}%
\bibitem [{\citenamefont {Chance}\ \emph {et~al.}(1978)\citenamefont {Chance},
  \citenamefont {Prock},\ and\ \citenamefont {Silbey}}]{chance1978}%
  \BibitemOpen
  \bibfield  {author} {\bibinfo {author} {\bibfnamefont {R.~R.}\ \bibnamefont
  {Chance}}, \bibinfo {author} {\bibfnamefont {A.}~\bibnamefont {Prock}}, \
  and\ \bibinfo {author} {\bibfnamefont {R.}~\bibnamefont {Silbey}},\
  }\bibfield  {title} {\enquote {\bibinfo {title} {Molecular fluorescence and
  energy transfer near interfaces},}\ }\href
  {http://books.google.com/books?hl=en&lr=&id=HlPB-WopREwC&oi=fnd&pg=PA1&dq=%22nature+of+the+states+of+the+metal+that+accept+this+energy%22+%22molecule.%E2%80%9D-+l+9+These+modes+have+become+of+interest+in%22+%22Section+I1+the+theory+of+a+dipole+antenna+near+a+metal+surface+is+given%22+&ots=kC3Av9B5Jc&sig=ZAR8DBGlrli0cV62Kky-6b0FN8M}
  {\bibfield  {journal} {\bibinfo  {journal} {Adv. Chem. Phys}\ }\textbf
  {\bibinfo {volume} {37}},\ \bibinfo {pages} {65} (\bibinfo {year}
  {1978})}\BibitemShut {NoStop}%
\bibitem [{\citenamefont {Swathi}\ and\ \citenamefont
  {Sebastian}(2009)}]{swathi2009}%
  \BibitemOpen
  \bibfield  {author} {\bibinfo {author} {\bibfnamefont {R.~S.}\ \bibnamefont
  {Swathi}}\ and\ \bibinfo {author} {\bibfnamefont {K.~L.}\ \bibnamefont
  {Sebastian}},\ }\bibfield  {title} {\enquote {\bibinfo {title} {Long range
  resonance energy transfer from a dye molecule to graphene has (distance)âˆ’4
  dependence},}\ }\href {\doibase 10.1063/1.3077292} {\bibfield  {journal}
  {\bibinfo  {journal} {The Journal of Chemical Physics}\ }\textbf {\bibinfo
  {volume} {130}},\ \bibinfo {pages} {086101} (\bibinfo {year}
  {2009})}\BibitemShut {NoStop}%
\bibitem [{\citenamefont {Basko}\ \emph {et~al.}(2000)\citenamefont {Basko},
  \citenamefont {Agranovich}, \citenamefont {Bassani},\ and\ \citenamefont
  {Rocca}}]{basko2000}%
  \BibitemOpen
  \bibfield  {author} {\bibinfo {author} {\bibfnamefont {D.~M.}\ \bibnamefont
  {Basko}}, \bibinfo {author} {\bibfnamefont {V.~M.}\ \bibnamefont
  {Agranovich}}, \bibinfo {author} {\bibfnamefont {F.}~\bibnamefont {Bassani}},
  \ and\ \bibinfo {author} {\bibfnamefont {G.~C.~La}\ \bibnamefont {Rocca}},\
  }\bibfield  {title} {\enquote {\bibinfo {title} {Energy transfer from a
  semiconductor quantum dot to an organic matrix},}\ }\href {\doibase
  10.1007/s100510050082} {\bibfield  {journal} {\bibinfo  {journal} {Eur. Phys.
  J. B}\ }\textbf {\bibinfo {volume} {13}},\ \bibinfo {pages} {653--659}
  (\bibinfo {year} {2000})}\BibitemShut {NoStop}%
\bibitem [{\citenamefont {Kos}\ \emph {et~al.}(2005)\citenamefont {Kos},
  \citenamefont {Achermann}, \citenamefont {Klimov},\ and\ \citenamefont
  {Smith}}]{kos2005}%
  \BibitemOpen
  \bibfield  {author} {\bibinfo {author} {\bibfnamefont {S.}~\bibnamefont
  {Kos}}, \bibinfo {author} {\bibfnamefont {M.}~\bibnamefont {Achermann}},
  \bibinfo {author} {\bibfnamefont {V.~I.}\ \bibnamefont {Klimov}}, \ and\
  \bibinfo {author} {\bibfnamefont {D.~L.}\ \bibnamefont {Smith}},\ }\bibfield
  {title} {\enquote {\bibinfo {title} {Different regimes of f\"orster-type
  energy transfer between an epitaxial quantum well and a proximal monolayer of
  semiconductor nanocrystals},}\ }\href {\doibase 10.1103/PhysRevB.71.205309}
  {\bibfield  {journal} {\bibinfo  {journal} {Phys. Rev. B}\ }\textbf {\bibinfo
  {volume} {71}},\ \bibinfo {pages} {205309} (\bibinfo {year}
  {2005})}\BibitemShut {NoStop}%
\bibitem [{\citenamefont {Mohamed}\ \emph {et~al.}(2005)\citenamefont
  {Mohamed}, \citenamefont {Tonti}, \citenamefont {Al-Salman}, \citenamefont
  {Chemseddine},\ and\ \citenamefont {Chergui}}]{mohamed2005}%
  \BibitemOpen
  \bibfield  {author} {\bibinfo {author} {\bibfnamefont {Mona~B.}\ \bibnamefont
  {Mohamed}}, \bibinfo {author} {\bibfnamefont {Dino}\ \bibnamefont {Tonti}},
  \bibinfo {author} {\bibfnamefont {Awos}\ \bibnamefont {Al-Salman}}, \bibinfo
  {author} {\bibfnamefont {Abdelkrim}\ \bibnamefont {Chemseddine}}, \ and\
  \bibinfo {author} {\bibfnamefont {Majed}\ \bibnamefont {Chergui}},\
  }\bibfield  {title} {\enquote {\bibinfo {title} {Synthesis of high quality
  zinc blende cdse nanocrystals},}\ }\href {\doibase 10.1021/jp051123e}
  {\bibfield  {journal} {\bibinfo  {journal} {The Journal of Physical Chemistry
  B}\ }\textbf {\bibinfo {volume} {109}},\ \bibinfo {pages} {10533--10537}
  (\bibinfo {year} {2005})}\BibitemShut {NoStop}%
\bibitem [{\citenamefont {Mahler}\ \emph {et~al.}(2008)\citenamefont {Mahler},
  \citenamefont {Spinicelli}, \citenamefont {Buil}, \citenamefont {Quelin},
  \citenamefont {Hermier},\ and\ \citenamefont {Dubertret}}]{mahler2008}%
  \BibitemOpen
  \bibfield  {author} {\bibinfo {author} {\bibfnamefont {Benoit}\ \bibnamefont
  {Mahler}}, \bibinfo {author} {\bibfnamefont {Piernicola}\ \bibnamefont
  {Spinicelli}}, \bibinfo {author} {\bibfnamefont {St\'ephanie}\ \bibnamefont
  {Buil}}, \bibinfo {author} {\bibfnamefont {Xavier}\ \bibnamefont {Quelin}},
  \bibinfo {author} {\bibfnamefont {Jean-Pierre}\ \bibnamefont {Hermier}}, \
  and\ \bibinfo {author} {\bibfnamefont {Benoit}\ \bibnamefont {Dubertret}},\
  }\bibfield  {title} {\enquote {\bibinfo {title} {Towards non-blinking
  colloidal quantum dots},}\ }\href {\doibase 10.1038/nmat2222} {\bibfield
  {journal} {\bibinfo  {journal} {Nature Materials}\ }\textbf {\bibinfo
  {volume} {7}},\ \bibinfo {pages} {659--664} (\bibinfo {year}
  {2008})}\BibitemShut {NoStop}%
\bibitem [{\citenamefont {Mahler}\ \emph {et~al.}(2012)\citenamefont {Mahler},
  \citenamefont {Nadal}, \citenamefont {Bouet}, \citenamefont {Patriarche},\
  and\ \citenamefont {Dubertret}}]{mahler2012}%
  \BibitemOpen
  \bibfield  {author} {\bibinfo {author} {\bibfnamefont {Benoit}\ \bibnamefont
  {Mahler}}, \bibinfo {author} {\bibfnamefont {Brice}\ \bibnamefont {Nadal}},
  \bibinfo {author} {\bibfnamefont {Cecile}\ \bibnamefont {Bouet}}, \bibinfo
  {author} {\bibfnamefont {Gilles}\ \bibnamefont {Patriarche}}, \ and\ \bibinfo
  {author} {\bibfnamefont {Benoit}\ \bibnamefont {Dubertret}},\ }\bibfield
  {title} {\enquote {\bibinfo {title} {{Core/Shell} colloidal semiconductor
  nanoplatelets},}\ }\href {\doibase 10.1021/ja307944d} {\bibfield  {journal}
  {\bibinfo  {journal} {Journal of the American Chemical Society}\ }\textbf
  {\bibinfo {volume} {134}},\ \bibinfo {pages} {18591--18598} (\bibinfo {year}
  {2012})}\BibitemShut {NoStop}%
\bibitem [{\citenamefont {Tessier}\ \emph {et~al.}(2013)\citenamefont
  {Tessier}, \citenamefont {Mahler}, \citenamefont {Nadal}, \citenamefont
  {Heuclin}, \citenamefont {Pedetti},\ and\ \citenamefont
  {Dubertret}}]{tessier2013}%
  \BibitemOpen
  \bibfield  {author} {\bibinfo {author} {\bibfnamefont {M.~D.}\ \bibnamefont
  {Tessier}}, \bibinfo {author} {\bibfnamefont {B.}~\bibnamefont {Mahler}},
  \bibinfo {author} {\bibfnamefont {B.}~\bibnamefont {Nadal}}, \bibinfo
  {author} {\bibfnamefont {H.}~\bibnamefont {Heuclin}}, \bibinfo {author}
  {\bibfnamefont {S.}~\bibnamefont {Pedetti}}, \ and\ \bibinfo {author}
  {\bibfnamefont {B.}~\bibnamefont {Dubertret}},\ }\bibfield  {title} {\enquote
  {\bibinfo {title} {Spectroscopy of colloidal semiconductor {Core/Shell}
  nanoplatelets with high quantum yield},}\ }\href {\doibase 10.1021/nl401538n}
  {\bibfield  {journal} {\bibinfo  {journal} {Nano Letters}\ }\textbf {\bibinfo
  {volume} {13}},\ \bibinfo {pages} {3321--3328} (\bibinfo {year}
  {2013})}\BibitemShut {NoStop}%
\bibitem [{\citenamefont {Hines}\ and\ \citenamefont
  {Guyot-Sionnest}(1996)}]{hines1996}%
  \BibitemOpen
  \bibfield  {author} {\bibinfo {author} {\bibfnamefont {Margaret~A.}\
  \bibnamefont {Hines}}\ and\ \bibinfo {author} {\bibfnamefont {Philippe}\
  \bibnamefont {Guyot-Sionnest}},\ }\bibfield  {title} {\enquote {\bibinfo
  {title} {Synthesis and characterization of strongly luminescing zns-capped
  cdse nanocrystals},}\ }\href {\doibase 10.1021/jp9530562} {\bibfield
  {journal} {\bibinfo  {journal} {The Journal of Physical Chemistry}\ }\textbf
  {\bibinfo {volume} {100}},\ \bibinfo {pages} {468--471} (\bibinfo {year}
  {1996})}\BibitemShut {NoStop}%
\bibitem [{\citenamefont {Li}\ \emph {et~al.}(2009)\citenamefont {Li},
  \citenamefont {Cai}, \citenamefont {An}, \citenamefont {Kim}, \citenamefont
  {Nah}, \citenamefont {Yang}, \citenamefont {Piner}, \citenamefont
  {Velamakanni}, \citenamefont {Jung}, \citenamefont {Tutuc}, \citenamefont
  {Banerjee}, \citenamefont {Colombo},\ and\ \citenamefont {Ruoff}}]{li2009}%
  \BibitemOpen
  \bibfield  {author} {\bibinfo {author} {\bibfnamefont {Xuesong}\ \bibnamefont
  {Li}}, \bibinfo {author} {\bibfnamefont {Weiwei}\ \bibnamefont {Cai}},
  \bibinfo {author} {\bibfnamefont {Jinho}\ \bibnamefont {An}}, \bibinfo
  {author} {\bibfnamefont {Seyoung}\ \bibnamefont {Kim}}, \bibinfo {author}
  {\bibfnamefont {Junghyo}\ \bibnamefont {Nah}}, \bibinfo {author}
  {\bibfnamefont {Dongxing}\ \bibnamefont {Yang}}, \bibinfo {author}
  {\bibfnamefont {Richard}\ \bibnamefont {Piner}}, \bibinfo {author}
  {\bibfnamefont {Aruna}\ \bibnamefont {Velamakanni}}, \bibinfo {author}
  {\bibfnamefont {Inhwa}\ \bibnamefont {Jung}}, \bibinfo {author}
  {\bibfnamefont {Emanuel}\ \bibnamefont {Tutuc}}, \bibinfo {author}
  {\bibfnamefont {Sanjay~K.}\ \bibnamefont {Banerjee}}, \bibinfo {author}
  {\bibfnamefont {Luigi}\ \bibnamefont {Colombo}}, \ and\ \bibinfo {author}
  {\bibfnamefont {Rodney~S.}\ \bibnamefont {Ruoff}},\ }\bibfield  {title}
  {\enquote {\bibinfo {title} {Large-area synthesis of high-quality and uniform
  graphene films on copper foils},}\ }\href {\doibase 10.1126/science.1171245}
  {\bibfield  {journal} {\bibinfo  {journal} {Science}\ }\textbf {\bibinfo
  {volume} {324}},\ \bibinfo {pages} {1312--1314} (\bibinfo {year}
  {2009})}\BibitemShut {NoStop}%
\bibitem [{\citenamefont {Wang}\ \emph
  {et~al.}(2008{\natexlab{b}})\citenamefont {Wang}, \citenamefont {Han},
  \citenamefont {Pi}, \citenamefont {McCreary}, \citenamefont {Miao},
  \citenamefont {Bao}, \citenamefont {Lau},\ and\ \citenamefont
  {Kawakami}}]{wang2008b}%
  \BibitemOpen
  \bibfield  {author} {\bibinfo {author} {\bibfnamefont {W.~H.}\ \bibnamefont
  {Wang}}, \bibinfo {author} {\bibfnamefont {W.}~\bibnamefont {Han}}, \bibinfo
  {author} {\bibfnamefont {K.}~\bibnamefont {Pi}}, \bibinfo {author}
  {\bibfnamefont {K.~M.}\ \bibnamefont {McCreary}}, \bibinfo {author}
  {\bibfnamefont {F.}~\bibnamefont {Miao}}, \bibinfo {author} {\bibfnamefont
  {W.}~\bibnamefont {Bao}}, \bibinfo {author} {\bibfnamefont {C.~N.}\
  \bibnamefont {Lau}}, \ and\ \bibinfo {author} {\bibfnamefont {R.~K.}\
  \bibnamefont {Kawakami}},\ }\bibfield  {title} {\enquote {\bibinfo {title}
  {Growth of atomically smooth mgo films on graphene by molecular beam
  epitaxy},}\ }\href {\doibase 10.1063/1.3013820} {\bibfield  {journal}
  {\bibinfo  {journal} {Applied Physics Letters}\ }\textbf {\bibinfo {volume}
  {93}},\ \bibinfo {eid} {183107} (\bibinfo {year}
  {2008}{\natexlab{b}})}\BibitemShut {NoStop}%
\bibitem [{\citenamefont {Leatherdale}\ \emph {et~al.}(2002)\citenamefont
  {Leatherdale}, \citenamefont {Woo}, \citenamefont {Mikulec},\ and\
  \citenamefont {Bawendi}}]{leatherdale2002}%
  \BibitemOpen
  \bibfield  {author} {\bibinfo {author} {\bibfnamefont {C.~A.}\ \bibnamefont
  {Leatherdale}}, \bibinfo {author} {\bibfnamefont {W.-K.}\ \bibnamefont
  {Woo}}, \bibinfo {author} {\bibfnamefont {F.~V.}\ \bibnamefont {Mikulec}}, \
  and\ \bibinfo {author} {\bibfnamefont {M.~G.}\ \bibnamefont {Bawendi}},\
  }\bibfield  {title} {\enquote {\bibinfo {title} {On the absorption cross
  section of {CdSe} nanocrystal quantum dots},}\ }\href {\doibase
  10.1021/jp025698c} {\bibfield  {journal} {\bibinfo  {journal} {The Journal of
  Physical Chemistry B}\ }\textbf {\bibinfo {volume} {106}},\ \bibinfo {pages}
  {7619--7622} (\bibinfo {year} {2002})}\BibitemShut {NoStop}%
\bibitem [{\citenamefont {Park}\ \emph {et~al.}(2011)\citenamefont {Park},
  \citenamefont {Malko}, \citenamefont {Vela}, \citenamefont {Chen},
  \citenamefont {Ghosh}, \citenamefont {Garcia-Santamaria}, \citenamefont
  {Hollingsworth}, \citenamefont {Klimov},\ and\ \citenamefont
  {Htoon}}]{park2011}%
  \BibitemOpen
  \bibfield  {author} {\bibinfo {author} {\bibfnamefont {Y.-S.}\ \bibnamefont
  {Park}}, \bibinfo {author} {\bibfnamefont {A.~V.}\ \bibnamefont {Malko}},
  \bibinfo {author} {\bibfnamefont {J.}~\bibnamefont {Vela}}, \bibinfo {author}
  {\bibfnamefont {Y.}~\bibnamefont {Chen}}, \bibinfo {author} {\bibfnamefont
  {Y.}~\bibnamefont {Ghosh}}, \bibinfo {author} {\bibfnamefont
  {F.}~\bibnamefont {Garcia-Santamaria}}, \bibinfo {author} {\bibfnamefont
  {J.~A.}\ \bibnamefont {Hollingsworth}}, \bibinfo {author} {\bibfnamefont
  {V.~I.}\ \bibnamefont {Klimov}}, \ and\ \bibinfo {author} {\bibfnamefont
  {H.}~\bibnamefont {Htoon}},\ }\bibfield  {title} {\enquote {\bibinfo {title}
  {Near-unity quantum yields of biexciton emission from
  $\mathrm{CdSe}/\mathrm{CdS}$ nanocrystals measured using single-particle
  spectroscopy},}\ }\href {\doibase 10.1103/PhysRevLett.106.187401} {\bibfield
  {journal} {\bibinfo  {journal} {Physical Review Letters}\ }\textbf {\bibinfo
  {volume} {106}},\ \bibinfo {pages} {187401} (\bibinfo {year}
  {2011})}\BibitemShut {NoStop}%
\bibitem [{\citenamefont {She}\ \emph {et~al.}(2014)\citenamefont {She},
  \citenamefont {Fedin}, \citenamefont {Dolzhnikov}, \citenamefont
  {DemortiÃ¨re}, \citenamefont {Schaller}, \citenamefont {Pelton},\ and\
  \citenamefont {Talapin}}]{she2014}%
  \BibitemOpen
  \bibfield  {author} {\bibinfo {author} {\bibfnamefont {Chunxing}\
  \bibnamefont {She}}, \bibinfo {author} {\bibfnamefont {Igor}\ \bibnamefont
  {Fedin}}, \bibinfo {author} {\bibfnamefont {Dmitriy~S.}\ \bibnamefont
  {Dolzhnikov}}, \bibinfo {author} {\bibfnamefont {Arnaud}\ \bibnamefont
  {DemortiÃ¨re}}, \bibinfo {author} {\bibfnamefont {Richard~D.}\ \bibnamefont
  {Schaller}}, \bibinfo {author} {\bibfnamefont {Matthew}\ \bibnamefont
  {Pelton}}, \ and\ \bibinfo {author} {\bibfnamefont {Dmitri~V.}\ \bibnamefont
  {Talapin}},\ }\bibfield  {title} {\enquote {\bibinfo {title} {Low-threshold
  stimulated emission using colloidal quantum wells},}\ }\href {\doibase
  10.1021/nl500775p} {\bibfield  {journal} {\bibinfo  {journal} {Nano Letters}\
  }\textbf {\bibinfo {volume} {14}},\ \bibinfo {pages} {2772--2777} (\bibinfo
  {year} {2014})}\BibitemShut {NoStop}%
\bibitem [{\citenamefont {Cichos}\ \emph {et~al.}(2007)\citenamefont {Cichos},
  \citenamefont {Vonborczyskowski},\ and\ \citenamefont {Orrit}}]{cichos2007}%
  \BibitemOpen
  \bibfield  {author} {\bibinfo {author} {\bibfnamefont {F}~\bibnamefont
  {Cichos}}, \bibinfo {author} {\bibfnamefont {C}~\bibnamefont
  {Vonborczyskowski}}, \ and\ \bibinfo {author} {\bibfnamefont {M}~\bibnamefont
  {Orrit}},\ }\bibfield  {title} {\enquote {\bibinfo {title} {Power-law
  intermittency of single emitters},}\ }\href {\doibase
  10.1016/j.cocis.2007.07.012} {\bibfield  {journal} {\bibinfo  {journal}
  {Current Opinion in Colloid \& Interface Science}\ }\textbf {\bibinfo
  {volume} {12}},\ \bibinfo {pages} {272--284} (\bibinfo {year}
  {2007})}\BibitemShut {NoStop}%
\bibitem [{\citenamefont {Spinicelli}\ \emph {et~al.}(2009)\citenamefont
  {Spinicelli}, \citenamefont {Buil}, \citenamefont {Qu\'elin}, \citenamefont
  {Mahler}, \citenamefont {Dubertret},\ and\ \citenamefont
  {Hermier}}]{spinicelli2009}%
  \BibitemOpen
  \bibfield  {author} {\bibinfo {author} {\bibfnamefont {P.}~\bibnamefont
  {Spinicelli}}, \bibinfo {author} {\bibfnamefont {S.}~\bibnamefont {Buil}},
  \bibinfo {author} {\bibfnamefont {X.}~\bibnamefont {Qu\'elin}}, \bibinfo
  {author} {\bibfnamefont {B.}~\bibnamefont {Mahler}}, \bibinfo {author}
  {\bibfnamefont {B.}~\bibnamefont {Dubertret}}, \ and\ \bibinfo {author}
  {\bibfnamefont {J.-P.}\ \bibnamefont {Hermier}},\ }\bibfield  {title}
  {\enquote {\bibinfo {title} {Bright and grey states in cdse-cds nanocrystals
  exhibiting strongly reduced blinking},}\ }\href {\doibase
  10.1103/PhysRevLett.102.136801} {\bibfield  {journal} {\bibinfo  {journal}
  {Physical Review Letters}\ }\textbf {\bibinfo {volume} {102}},\ \bibinfo
  {pages} {136801} (\bibinfo {year} {2009})}\BibitemShut {NoStop}%
\bibitem [{\citenamefont {Malko}\ \emph {et~al.}(2011)\citenamefont {Malko},
  \citenamefont {Park}, \citenamefont {Sampat}, \citenamefont {Galland},
  \citenamefont {Vela}, \citenamefont {Chen}, \citenamefont {Hollingsworth},
  \citenamefont {Klimov},\ and\ \citenamefont {Htoon}}]{malko2012}%
  \BibitemOpen
  \bibfield  {author} {\bibinfo {author} {\bibfnamefont {Anton~V.}\
  \bibnamefont {Malko}}, \bibinfo {author} {\bibfnamefont {Young-Shin}\
  \bibnamefont {Park}}, \bibinfo {author} {\bibfnamefont {Siddharth}\
  \bibnamefont {Sampat}}, \bibinfo {author} {\bibfnamefont {Christophe}\
  \bibnamefont {Galland}}, \bibinfo {author} {\bibfnamefont {Javier}\
  \bibnamefont {Vela}}, \bibinfo {author} {\bibfnamefont {Yongfen}\
  \bibnamefont {Chen}}, \bibinfo {author} {\bibfnamefont {Jennifer~A.}\
  \bibnamefont {Hollingsworth}}, \bibinfo {author} {\bibfnamefont {Victor~I.}\
  \bibnamefont {Klimov}}, \ and\ \bibinfo {author} {\bibfnamefont {Han}\
  \bibnamefont {Htoon}},\ }\bibfield  {title} {\enquote {\bibinfo {title}
  {Pump-intensity- and shell-thickness-dependent evolution of photoluminescence
  blinking in individual {Core/Shell} {CdSe/CdS} nanocrystals},}\ }\href
  {\doibase 10.1021/nl2025272} {\bibfield  {journal} {\bibinfo  {journal} {Nano
  Letters}\ }\textbf {\bibinfo {volume} {11}},\ \bibinfo {pages} {5213--5218}
  (\bibinfo {year} {2011})}\BibitemShut {NoStop}%
\bibitem [{\citenamefont {Tessier}\ \emph {et~al.}(2012)\citenamefont
  {Tessier}, \citenamefont {Javaux}, \citenamefont {Maksimovic}, \citenamefont
  {Loriette},\ and\ \citenamefont {Dubertret}}]{tessier2012}%
  \BibitemOpen
  \bibfield  {author} {\bibinfo {author} {\bibfnamefont {Micka\"el~D.}\
  \bibnamefont {Tessier}}, \bibinfo {author} {\bibfnamefont {Cl\'ementine}\
  \bibnamefont {Javaux}}, \bibinfo {author} {\bibfnamefont {Ivan}\ \bibnamefont
  {Maksimovic}}, \bibinfo {author} {\bibfnamefont {Vincent}\ \bibnamefont
  {Loriette}}, \ and\ \bibinfo {author} {\bibfnamefont {Benoit}\ \bibnamefont
  {Dubertret}},\ }\bibfield  {title} {\enquote {\bibinfo {title} {Spectroscopy
  of single {CdSe} nanoplatelets},}\ }\href {\doibase 10.1021/nn3014855}
  {\bibfield  {journal} {\bibinfo  {journal} {{ACS} Nano}\ }\textbf {\bibinfo
  {volume} {6}},\ \bibinfo {pages} {6751--6758} (\bibinfo {year}
  {2012})}\BibitemShut {NoStop}%
\bibitem [{\citenamefont {Swathi}\ and\ \citenamefont
  {Sebastian}(2008)}]{swathi2008}%
  \BibitemOpen
  \bibfield  {author} {\bibinfo {author} {\bibfnamefont {R.~S.}\ \bibnamefont
  {Swathi}}\ and\ \bibinfo {author} {\bibfnamefont {K.~L.}\ \bibnamefont
  {Sebastian}},\ }\bibfield  {title} {\enquote {\bibinfo {title} {Resonance
  energy transfer from a dye molecule to graphene},}\ }\href {\doibase
  10.1063/1.2956498} {\bibfield  {journal} {\bibinfo  {journal} {The Journal of
  Chemical Physics}\ }\textbf {\bibinfo {volume} {129}},\ \bibinfo {pages}
  {054703} (\bibinfo {year} {2008})}\BibitemShut {NoStop}%
\bibitem [{\citenamefont {Malic}\ \emph {et~al.}(2014)\citenamefont {Malic},
  \citenamefont {Appel}, \citenamefont {Hofmann},\ and\ \citenamefont
  {Rubio}}]{malic2014}%
  \BibitemOpen
  \bibfield  {author} {\bibinfo {author} {\bibfnamefont {Ermin}\ \bibnamefont
  {Malic}}, \bibinfo {author} {\bibfnamefont {Heiko}\ \bibnamefont {Appel}},
  \bibinfo {author} {\bibfnamefont {Oliver~T.}\ \bibnamefont {Hofmann}}, \ and\
  \bibinfo {author} {\bibfnamefont {Angel}\ \bibnamefont {Rubio}},\ }\bibfield
  {title} {\enquote {\bibinfo {title} {F\"orster-induced energy transfer in
  functionalized graphene},}\ }\href {\doibase 10.1021/jp5019636} {\bibfield
  {journal} {\bibinfo  {journal} {The Journal of Physical Chemistry C}\
  }\textbf {\bibinfo {volume} {118}},\ \bibinfo {pages} {9283--9289} (\bibinfo
  {year} {2014})}\BibitemShut {NoStop}%
\bibitem [{\citenamefont {Califano}\ \emph {et~al.}(2005)\citenamefont
  {Califano}, \citenamefont {Franceschetti},\ and\ \citenamefont
  {Zunger}}]{califano2005}%
  \BibitemOpen
  \bibfield  {author} {\bibinfo {author} {\bibfnamefont {Marco}\ \bibnamefont
  {Califano}}, \bibinfo {author} {\bibfnamefont {Alberto}\ \bibnamefont
  {Franceschetti}}, \ and\ \bibinfo {author} {\bibfnamefont {Alex}\
  \bibnamefont {Zunger}},\ }\bibfield  {title} {\enquote {\bibinfo {title}
  {Temperature dependence of excitonic radiative decay in {CdSe} quantum dots:
  The role of surface hole traps},}\ }\href {\doibase 10.1021/nl051027p}
  {\bibfield  {journal} {\bibinfo  {journal} {Nano Letters}\ }\textbf {\bibinfo
  {volume} {5}},\ \bibinfo {pages} {2360--2364} (\bibinfo {year}
  {2005})}\BibitemShut {NoStop}%
\bibitem [{\citenamefont {Blachnik}\ \emph {et~al.}(1999)\citenamefont
  {Blachnik}, \citenamefont {Chu}, \citenamefont {Galazka}, \citenamefont
  {Geurts}, \citenamefont {Gutowski}, \citenamefont {H\"onerlage},
  \citenamefont {Hofmann}, \citenamefont {Kossut}, \citenamefont {L\'evy},
  \citenamefont {Michler}, \citenamefont {Neukirch}, \citenamefont {Story},
  \citenamefont {Strauch},\ and\ \citenamefont {Waag}}]{LandoltBornstein}%
  \BibitemOpen
  \bibfield  {author} {\bibinfo {author} {\bibfnamefont {R.}~\bibnamefont
  {Blachnik}}, \bibinfo {author} {\bibfnamefont {J.}~\bibnamefont {Chu}},
  \bibinfo {author} {\bibfnamefont {R.R.}\ \bibnamefont {Galazka}}, \bibinfo
  {author} {\bibfnamefont {J.}~\bibnamefont {Geurts}}, \bibinfo {author}
  {\bibfnamefont {J.}~\bibnamefont {Gutowski}}, \bibinfo {author}
  {\bibfnamefont {B.}~\bibnamefont {H\"onerlage}}, \bibinfo {author}
  {\bibfnamefont {D.}~\bibnamefont {Hofmann}}, \bibinfo {author} {\bibfnamefont
  {J.}~\bibnamefont {Kossut}}, \bibinfo {author} {\bibfnamefont
  {R.}~\bibnamefont {L\'evy}}, \bibinfo {author} {\bibfnamefont
  {P.}~\bibnamefont {Michler}}, \bibinfo {author} {\bibfnamefont
  {U.}~\bibnamefont {Neukirch}}, \bibinfo {author} {\bibfnamefont
  {T.}~\bibnamefont {Story}}, \bibinfo {author} {\bibfnamefont
  {D.}~\bibnamefont {Strauch}}, \ and\ \bibinfo {author} {\bibfnamefont
  {A.}~\bibnamefont {Waag}},\ }\bibfield  {title} {\enquote {\bibinfo {title}
  {Semiconductors: Ii-vi and i-vii compounds; semimagnetic compounds},}\ }in\
  \href {\doibase 10.1007/b71137} {\emph {\bibinfo {booktitle}
  {Landolt-B\"ornstein, New Series III/41B}}},\ \bibinfo {editor} {edited by\
  \bibinfo {editor} {\bibfnamefont {U.}~\bibnamefont {R\"ossler}}}\ (\bibinfo
  {publisher} {Springer Verlag},\ \bibinfo {year} {1999})\BibitemShut {NoStop}%
\bibitem [{\citenamefont {Johannsen}\ \emph {et~al.}(2013)\citenamefont
  {Johannsen}, \citenamefont {Ulstrup}, \citenamefont {Cilento}, \citenamefont
  {Crepaldi}, \citenamefont {Zacchigna}, \citenamefont {Cacho}, \citenamefont
  {Turcu}, \citenamefont {Springate}, \citenamefont {Fromm}, \citenamefont
  {Raidel}, \citenamefont {Seyller}, \citenamefont {Parmigiani}, \citenamefont
  {Grioni},\ and\ \citenamefont {Hofmann}}]{johannsen2013}%
  \BibitemOpen
  \bibfield  {author} {\bibinfo {author} {\bibfnamefont {Jens~Christian}\
  \bibnamefont {Johannsen}}, \bibinfo {author} {\bibfnamefont {S\o{}ren}\
  \bibnamefont {Ulstrup}}, \bibinfo {author} {\bibfnamefont {Federico}\
  \bibnamefont {Cilento}}, \bibinfo {author} {\bibfnamefont {Alberto}\
  \bibnamefont {Crepaldi}}, \bibinfo {author} {\bibfnamefont {Michele}\
  \bibnamefont {Zacchigna}}, \bibinfo {author} {\bibfnamefont {Cephise}\
  \bibnamefont {Cacho}}, \bibinfo {author} {\bibfnamefont {I.~C.~Edmond}\
  \bibnamefont {Turcu}}, \bibinfo {author} {\bibfnamefont {Emma}\ \bibnamefont
  {Springate}}, \bibinfo {author} {\bibfnamefont {Felix}\ \bibnamefont
  {Fromm}}, \bibinfo {author} {\bibfnamefont {Christian}\ \bibnamefont
  {Raidel}}, \bibinfo {author} {\bibfnamefont {Thomas}\ \bibnamefont
  {Seyller}}, \bibinfo {author} {\bibfnamefont {Fulvio}\ \bibnamefont
  {Parmigiani}}, \bibinfo {author} {\bibfnamefont {Marco}\ \bibnamefont
  {Grioni}}, \ and\ \bibinfo {author} {\bibfnamefont {Philip}\ \bibnamefont
  {Hofmann}},\ }\bibfield  {title} {\enquote {\bibinfo {title} {Direct view of
  hot carrier dynamics in graphene},}\ }\href {\doibase
  10.1103/PhysRevLett.111.027403} {\bibfield  {journal} {\bibinfo  {journal}
  {Physical Review Letters}\ }\textbf {\bibinfo {volume} {111}},\ \bibinfo
  {pages} {027403} (\bibinfo {year} {2013})}\BibitemShut {NoStop}%
\bibitem [{\citenamefont {Gierz}\ \emph {et~al.}(2013)\citenamefont {Gierz},
  \citenamefont {Petersen}, \citenamefont {Mitrano}, \citenamefont {Cacho},
  \citenamefont {Turcu}, \citenamefont {Springate}, \citenamefont {St\"ohr},
  \citenamefont {K\"ohler}, \citenamefont {Starke},\ and\ \citenamefont
  {Cavalleri}}]{gierz2013}%
  \BibitemOpen
  \bibfield  {author} {\bibinfo {author} {\bibfnamefont {Isabella}\
  \bibnamefont {Gierz}}, \bibinfo {author} {\bibfnamefont {Jesse~C.}\
  \bibnamefont {Petersen}}, \bibinfo {author} {\bibfnamefont {Matteo}\
  \bibnamefont {Mitrano}}, \bibinfo {author} {\bibfnamefont {Cephise}\
  \bibnamefont {Cacho}}, \bibinfo {author} {\bibfnamefont {I.~C.~Edmond}\
  \bibnamefont {Turcu}}, \bibinfo {author} {\bibfnamefont {Emma}\ \bibnamefont
  {Springate}}, \bibinfo {author} {\bibfnamefont {Alexander}\ \bibnamefont
  {St\"ohr}}, \bibinfo {author} {\bibfnamefont {Axel}\ \bibnamefont
  {K\"ohler}}, \bibinfo {author} {\bibfnamefont {Ulrich}\ \bibnamefont
  {Starke}}, \ and\ \bibinfo {author} {\bibfnamefont {Andrea}\ \bibnamefont
  {Cavalleri}},\ }\bibfield  {title} {\enquote {\bibinfo {title} {Snapshots of
  non-equilibrium dirac carrier distributions in graphene},}\ }\href {\doibase
  10.1038/nmat3757} {\bibfield  {journal} {\bibinfo  {journal} {Nature
  Materials}\ }\textbf {\bibinfo {volume} {12}},\ \bibinfo {pages} {1119--1124}
  (\bibinfo {year} {2013})}\BibitemShut {NoStop}%
\end{thebibliography}

\begin{thebibliography}{8}%
\makeatletter
\providecommand \@ifxundefined [1]{%
 \@ifx{#1\undefined}
}%
\providecommand \@ifnum [1]{%
 \ifnum #1\expandafter \@firstoftwo
 \else \expandafter \@secondoftwo
 \fi
}%
\providecommand \@ifx [1]{%
 \ifx #1\expandafter \@firstoftwo
 \else \expandafter \@secondoftwo
 \fi
}%
\providecommand \natexlab [1]{#1}%
\providecommand \enquote  [1]{``#1''}%
\providecommand \bibnamefont  [1]{#1}%
\providecommand \bibfnamefont [1]{#1}%
\providecommand \citenamefont [1]{#1}%
\providecommand \href@noop [0]{\@secondoftwo}%
\providecommand \href [0]{\begingroup \@sanitize@url \@href}%
\providecommand \@href[1]{\@@startlink{#1}\@@href}%
\providecommand \@@href[1]{\endgroup#1\@@endlink}%
\providecommand \@sanitize@url [0]{\catcode `\\12\catcode `\$12\catcode
  `\&12\catcode `\#12\catcode `\^12\catcode `\_12\catcode `\%12\relax}%
\providecommand \@@startlink[1]{}%
\providecommand \@@endlink[0]{}%
\providecommand \url  [0]{\begingroup\@sanitize@url \@url }%
\providecommand \@url [1]{\endgroup\@href {#1}{\urlprefix }}%
\providecommand \urlprefix  [0]{URL }%
\providecommand \Eprint [0]{\href }%
\providecommand \doibase [0]{http://dx.doi.org/}%
\providecommand \selectlanguage [0]{\@gobble}%
\providecommand \bibinfo  [0]{\@secondoftwo}%
\providecommand \bibfield  [0]{\@secondoftwo}%
\providecommand \translation [1]{[#1]}%
\providecommand \BibitemOpen [0]{}%
\providecommand \bibitemStop [0]{}%
\providecommand \bibitemNoStop [0]{.\EOS\space}%
\providecommand \EOS [0]{\spacefactor3000\relax}%
\providecommand \BibitemShut  [1]{\csname bibitem#1\endcsname}%
\let\auto@bib@innerbib\@empty
\bibitem [{\citenamefont {Yang}\ \emph {et~al.}(2005)\citenamefont {Yang},
  \citenamefont {Wu}, \citenamefont {Williams},\ and\ \citenamefont
  {Cao}}]{yang2005}%
  \BibitemOpen
  \bibfield  {author} {\bibinfo {author} {\bibfnamefont {Yongan~Andrew}\
  \bibnamefont {Yang}}, \bibinfo {author} {\bibfnamefont {Huimeng}\
  \bibnamefont {Wu}}, \bibinfo {author} {\bibfnamefont {Kathryn~R.}\
  \bibnamefont {Williams}}, \ and\ \bibinfo {author} {\bibfnamefont
  {Y.~Charles}\ \bibnamefont {Cao}},\ }\bibfield  {title} {\enquote {\bibinfo
  {title} {Synthesis of cdse and cdte nanocrystals without precursor
  injection},}\ }\href {\doibase 10.1002/anie.200502279} {\bibfield  {journal}
  {\bibinfo  {journal} {Angewandte Chemie International Edition}\ }\textbf
  {\bibinfo {volume} {44}},\ \bibinfo {pages} {6712--6715} (\bibinfo {year}
  {2005})}\BibitemShut {NoStop}%
\bibitem [{\citenamefont {Mahler}\ \emph {et~al.}(2008)\citenamefont {Mahler},
  \citenamefont {Spinicelli}, \citenamefont {Buil}, \citenamefont {Quelin},
  \citenamefont {Hermier},\ and\ \citenamefont {Dubertret}}]{mahler2008}%
  \BibitemOpen
  \bibfield  {author} {\bibinfo {author} {\bibfnamefont {Benoit}\ \bibnamefont
  {Mahler}}, \bibinfo {author} {\bibfnamefont {Piernicola}\ \bibnamefont
  {Spinicelli}}, \bibinfo {author} {\bibfnamefont {St\'ephanie}\ \bibnamefont
  {Buil}}, \bibinfo {author} {\bibfnamefont {Xavier}\ \bibnamefont {Quelin}},
  \bibinfo {author} {\bibfnamefont {Jean-Pierre}\ \bibnamefont {Hermier}}, \
  and\ \bibinfo {author} {\bibfnamefont {Benoit}\ \bibnamefont {Dubertret}},\
  }\bibfield  {title} {\enquote {\bibinfo {title} {Towards non-blinking
  colloidal quantumdots},}\ }\href {\doibase 10.1038/nmat2222} {\bibfield
  {journal} {\bibinfo  {journal} {Nature Materials}\ }\textbf {\bibinfo
  {volume} {7}},\ \bibinfo {pages} {659--664} (\bibinfo {year}
  {2008})}\BibitemShut {NoStop}%
\bibitem [{\citenamefont {Mohamed}\ \emph {et~al.}(2005)\citenamefont
  {Mohamed}, \citenamefont {Tonti}, \citenamefont {Al-Salman}, \citenamefont
  {Chemseddine},\ and\ \citenamefont {Chergui}}]{mohamed2005}%
  \BibitemOpen
  \bibfield  {author} {\bibinfo {author} {\bibfnamefont {Mona~B.}\ \bibnamefont
  {Mohamed}}, \bibinfo {author} {\bibfnamefont {Dino}\ \bibnamefont {Tonti}},
  \bibinfo {author} {\bibfnamefont {Awos}\ \bibnamefont {Al-Salman}}, \bibinfo
  {author} {\bibfnamefont {Abdelkrim}\ \bibnamefont {Chemseddine}}, \ and\
  \bibinfo {author} {\bibfnamefont {Majed}\ \bibnamefont {Chergui}},\
  }\bibfield  {title} {\enquote {\bibinfo {title} {Synthesis of high quality
  zinc blende cdse nanocrystals},}\ }\href {\doibase 10.1021/jp051123e}
  {\bibfield  {journal} {\bibinfo  {journal} {The Journal of Physical Chemistry
  B}\ }\textbf {\bibinfo {volume} {109}},\ \bibinfo {pages} {10533--10537}
  (\bibinfo {year} {2005})}\BibitemShut {NoStop}%
\bibitem [{\citenamefont {Ithurria}\ and\ \citenamefont
  {Dubertret}(2008)}]{ithurria2008}%
  \BibitemOpen
  \bibfield  {author} {\bibinfo {author} {\bibfnamefont {Sandrine}\
  \bibnamefont {Ithurria}}\ and\ \bibinfo {author} {\bibfnamefont {Benoit}\
  \bibnamefont {Dubertret}},\ }\bibfield  {title} {\enquote {\bibinfo {title}
  {Quasi {2D} colloidal {CdSe} platelets with thicknesses controlled at the
  atomic level},}\ }\href {\doibase 10.1021/ja807724e} {\bibfield  {journal}
  {\bibinfo  {journal} {Journal of the American Chemical Society}\ }\textbf
  {\bibinfo {volume} {130}},\ \bibinfo {pages} {16504--16505} (\bibinfo {year}
  {2008})}\BibitemShut {NoStop}%
\bibitem [{\citenamefont {Li}\ \emph {et~al.}(2009)\citenamefont {Li},
  \citenamefont {Cai}, \citenamefont {An}, \citenamefont {Kim}, \citenamefont
  {Nah}, \citenamefont {Yang}, \citenamefont {Piner}, \citenamefont
  {Velamakanni}, \citenamefont {Jung}, \citenamefont {Tutuc}, \citenamefont
  {Banerjee}, \citenamefont {Colombo},\ and\ \citenamefont {Ruoff}}]{li2009}%
  \BibitemOpen
  \bibfield  {author} {\bibinfo {author} {\bibfnamefont {Xuesong}\ \bibnamefont
  {Li}}, \bibinfo {author} {\bibfnamefont {Weiwei}\ \bibnamefont {Cai}},
  \bibinfo {author} {\bibfnamefont {Jinho}\ \bibnamefont {An}}, \bibinfo
  {author} {\bibfnamefont {Seyoung}\ \bibnamefont {Kim}}, \bibinfo {author}
  {\bibfnamefont {Junghyo}\ \bibnamefont {Nah}}, \bibinfo {author}
  {\bibfnamefont {Dongxing}\ \bibnamefont {Yang}}, \bibinfo {author}
  {\bibfnamefont {Richard}\ \bibnamefont {Piner}}, \bibinfo {author}
  {\bibfnamefont {Aruna}\ \bibnamefont {Velamakanni}}, \bibinfo {author}
  {\bibfnamefont {Inhwa}\ \bibnamefont {Jung}}, \bibinfo {author}
  {\bibfnamefont {Emanuel}\ \bibnamefont {Tutuc}}, \bibinfo {author}
  {\bibfnamefont {Sanjay~K.}\ \bibnamefont {Banerjee}}, \bibinfo {author}
  {\bibfnamefont {Luigi}\ \bibnamefont {Colombo}}, \ and\ \bibinfo {author}
  {\bibfnamefont {Rodney~S.}\ \bibnamefont {Ruoff}},\ }\bibfield  {title}
  {\enquote {\bibinfo {title} {Large-area synthesis of high-quality and uniform
  graphene films on copper foils},}\ }\href {\doibase 10.1126/science.1171245}
  {\bibfield  {journal} {\bibinfo  {journal} {Science}\ }\textbf {\bibinfo
  {volume} {324}},\ \bibinfo {pages} {1312--1314} (\bibinfo {year}
  {2009})}\BibitemShut {NoStop}%
\bibitem [{\citenamefont {Godel}\ \emph {et~al.}(2013)\citenamefont {Godel},
  \citenamefont {Pichonat}, \citenamefont {Vignaud}, \citenamefont {Majjad},
  \citenamefont {Metten}, \citenamefont {Henry}, \citenamefont {Berciaud},
  \citenamefont {Dayen},\ and\ \citenamefont {Halley}}]{godel2013}%
  \BibitemOpen
  \bibfield  {author} {\bibinfo {author} {\bibfnamefont {Florian}\ \bibnamefont
  {Godel}}, \bibinfo {author} {\bibfnamefont {Emmanuelle}\ \bibnamefont
  {Pichonat}}, \bibinfo {author} {\bibfnamefont {Dominique}\ \bibnamefont
  {Vignaud}}, \bibinfo {author} {\bibfnamefont {Hicham}\ \bibnamefont
  {Majjad}}, \bibinfo {author} {\bibfnamefont {Dominik}\ \bibnamefont
  {Metten}}, \bibinfo {author} {\bibfnamefont {Yves}\ \bibnamefont {Henry}},
  \bibinfo {author} {\bibfnamefont {Stéphane}\ \bibnamefont {Berciaud}},
  \bibinfo {author} {\bibfnamefont {Jean-Francois}\ \bibnamefont {Dayen}}, \
  and\ \bibinfo {author} {\bibfnamefont {David}\ \bibnamefont {Halley}},\
  }\bibfield  {title} {\enquote {\bibinfo {title} {Epitaxy of {MgO} magnetic
  tunnel barriers on epitaxial graphene},}\ }\href {\doibase
  10.1088/0957-4484/24/47/475708} {\bibfield  {journal} {\bibinfo  {journal}
  {Nanotechnology}\ }\textbf {\bibinfo {volume} {24}},\ \bibinfo {pages}
  {475708} (\bibinfo {year} {2013})}\BibitemShut {NoStop}%
\bibitem [{\citenamefont {Ferrari}\ and\ \citenamefont
  {Basko}(2013)}]{ferrari2013}%
  \BibitemOpen
  \bibfield  {author} {\bibinfo {author} {\bibfnamefont {Andrea~C.}\
  \bibnamefont {Ferrari}}\ and\ \bibinfo {author} {\bibfnamefont {Denis~M.}\
  \bibnamefont {Basko}},\ }\bibfield  {title} {\enquote {\bibinfo {title}
  {Raman spectroscopy as a versatile tool for studying the properties of
  graphene},}\ }\href {\doibase 10.1038/nnano.2013.46} {\bibfield  {journal}
  {\bibinfo  {journal} {Nature Nanotechnology}\ }\textbf {\bibinfo {volume}
  {8}},\ \bibinfo {pages} {235--246} (\bibinfo {year} {2013})}\BibitemShut
  {NoStop}%
\bibitem [{\citenamefont {Das}\ \emph {et~al.}(2008)\citenamefont {Das},
  \citenamefont {Pisana}, \citenamefont {Chakraborty}, \citenamefont
  {Piscanec}, \citenamefont {Saha}, \citenamefont {Waghmare}, \citenamefont
  {Novoselov}, \citenamefont {Krishnamurthy}, \citenamefont {Geim},
  \citenamefont {Ferrari},\ and\ \citenamefont {Sood}}]{das2008}%
  \BibitemOpen
  \bibfield  {author} {\bibinfo {author} {\bibfnamefont {A.}~\bibnamefont
  {Das}}, \bibinfo {author} {\bibfnamefont {S.}~\bibnamefont {Pisana}},
  \bibinfo {author} {\bibfnamefont {B.}~\bibnamefont {Chakraborty}}, \bibinfo
  {author} {\bibfnamefont {S.}~\bibnamefont {Piscanec}}, \bibinfo {author}
  {\bibfnamefont {S.~K.}\ \bibnamefont {Saha}}, \bibinfo {author}
  {\bibfnamefont {U.~V.}\ \bibnamefont {Waghmare}}, \bibinfo {author}
  {\bibfnamefont {K.~S.}\ \bibnamefont {Novoselov}}, \bibinfo {author}
  {\bibfnamefont {H.~R.}\ \bibnamefont {Krishnamurthy}}, \bibinfo {author}
  {\bibfnamefont {A.~K.}\ \bibnamefont {Geim}}, \bibinfo {author}
  {\bibfnamefont {A.~C.}\ \bibnamefont {Ferrari}}, \ and\ \bibinfo {author}
  {\bibfnamefont {A.~K.}\ \bibnamefont {Sood}},\ }\bibfield  {title} {\enquote
  {\bibinfo {title} {Monitoring dopants by raman scattering in an
  electrochemically top-gated graphene transistor},}\ }\href {\doibase
  10.1038/nnano.2008.67} {\bibfield  {journal} {\bibinfo  {journal} {Nature
  Nanotechnology}\ }\textbf {\bibinfo {volume} {3}},\ \bibinfo {pages}
  {210--215} (\bibinfo {year} {2008})}\BibitemShut {NoStop}%
\end{thebibliography}

%


\end{document}